\documentclass{emulateapj}

\newcommand\hour{\mbox{$^{\mathrm h}$}}
\newcommand\minute{\mbox{$^{\mathrm m}$}}
\newcommand\second{\mbox{$^{\mathrm s}$}}
\newcommand\vect[1]{\mbox{\boldmath{$#1$}}}
\newcommand\tjptensor[1]{\mbox{{$\bf #1$}}}

\begin{document}

\journalinfo{Accepted for publication in The Astrophysical Journal}
\submitted{}

\title{The Anisotropy of the Microwave Background to $\ell=3500$: Mosaic Observations with the Cosmic Background Imager}
\shortauthors{PEARSON ET AL.}
\shorttitle{MOSAIC OBSERVATIONS WITH THE COSMIC BACKGROUND IMAGER}

\author{T. J. Pearson, B. S. Mason,\altaffilmark{1}  A. C. S. Readhead, M. C. Shepherd,  J. L. Sievers,\\
P. S. Udomprasert, J. K. Cartwright, A. J. Farmer, and S. Padin}

\affil{Owens Valley Radio Observatory, California Institute of Technology,
Pasadena, CA 91125}

\author{S. T. Myers}

\affil{National Radio Astronomy Observatory, P.O. Box O, Socorro, NM 87801}

\author{J. R. Bond, C. R. Contaldi, U.-L. Pen, and S. Prunet\altaffilmark{2}}

\affil{Canadian Institute for Theoretical Astrophysics, University of Toronto, 60 St. George Street, Toronto, Ontario, M5S 3H8, Canada}

\author{D. Pogosyan\altaffilmark{3}}

\affil{Department of Physics, University of Alberta, Edmonton, Alberta T6G 2J1, Canada}

\author{J. E. Carlstrom, J. Kovac, E. M. Leitch, and C. Pryke}

\affil{University of Chicago, 5640 South Ellis Ave., Chicago, IL 60637}

\author{N. W. Halverson and W. L. Holzapfel}

\affil{University of California, 366 LeConte Hall, Berkeley, CA 94720-7300}

\author{P. Altamirano, L. Bronfman, S. Casassus, and J. May}

\affil{Departamento de Astronom\'{\i}a, Universidad de Chile, Casilla 36-D, 
Santiago, 
Chile}

\and\author{M. Joy}

\affil{Dept. of Space Science, SD50, NASA Marshall Space Flight Center, 
Huntsville, AL 
35812}

\altaffiltext{1}{Currently at National Radio Astronomy Observatory, P.O. Box 2, Green Bank, WV 24944.}
\altaffiltext{2}{Currently at Institut d'Astrophysique de Paris, 98bis Boulevard Arago, F 75014 Paris, France.}
\altaffiltext{3}{Also at Canadian Institute for Theoretical Astrophysics.}

\begin{abstract}
Using the Cosmic Background Imager, a 13-element interferometer array
operating in the 26--36 GHz frequency band, we have observed 40
deg$^2$ of sky in three pairs of fields, each $\sim145\arcmin \times
165\arcmin$, using overlapping pointings (mosaicing). We present
images and power spectra of the cosmic microwave background radiation
in these mosaic fields.  We remove ground radiation and other
low-level contaminating signals by differencing matched observations
of the fields in each pair. The primary foreground contamination is
due to point sources (radio galaxies and quasars). We have subtracted
the strongest sources from the data using higher-resolution
measurements, and we have projected out the response to other sources
of known position in the power-spectrum analysis.  The images show
features on scales $\sim 6\arcmin$--$15\arcmin$, corresponding to masses
$\sim 5$--$80 \times 10^{14} M_\sun$ at the surface of last scattering, which
are likely to be the seeds of clusters of galaxies. The power
spectrum estimates have a resolution $\Delta \ell \approx 200$ and are
consistent with earlier results in the multipole range $\ell \lesssim
1000$. The power spectrum is detected with high signal-to-noise ratio
in the range $300 \lesssim \ell \lesssim 1700$. For $1700 \lesssim
\ell \lesssim 3000$ the observations are consistent with the results
from more sensitive CBI deep-field observations. The results agree
with the extrapolation of cosmological models fitted to observations
at lower $\ell$, and show the predicted drop at high $\ell$ (the
``damping tail'').
\end{abstract}


\section{Introduction}

The Cosmic Background Imager (CBI) is a 13-element radio
interferometer array designed to image the cosmic microwave background
radiation (CMB) and measure its angular power spectrum in the 26--36
GHz frequency band. The power of accurate measurements of the CMB
power spectrum to constrain cosmological models and obtain precise
estimates of critical cosmological parameters has been demonstrated by
many theoretical and observational studies. The most recent
experiments -- the BOOMERANG \citep{Netterfield02} and MAXIMA
\citep{Lee01} balloon-borne bolometers and the DASI interferometer
array at the South Pole \citep{Halverson02} -- have measured the power
spectrum at multipoles $\ell \lesssim 1000$ (angular scales $\gtrsim
20\arcmin$). The CBI has the potential to extend these measurements
to $\ell \sim 3500$.

This paper is the third in a series reporting results from the CBI.
Preliminary results were presented by \citet[hereafter Paper
I]{Padin01}. The accompanying paper \citep[hereafter Paper
II]{Mason02} presents our estimate of the power spectrum for $\ell
\lesssim 3500$ from observations of three pairs of $45\arcmin$ (FWHM)
deep fields made in our first observing season, 2000
January--December. The present paper (Paper III) presents
complementary results from first-season observations of three pairs of
mosaic fields each about $145\arcmin \times 165\arcmin$ (total $\approx
40$~deg$^2$), using the mosaicing method. These observations have
higher resolution in $\ell$ than those of Paper II, and they have
greater sensitivity at low $\ell$ owing to the reduced cosmic
variance, but they are less sensitive at high
$\ell$.  Further observations made in 2001, which are currently being
analyzed, will increase the sensitivity and improve the resolution of
our power spectrum estimate. The method we use for extracting power
spectrum estimates from interferometry data is described by
\citet[hereafter Paper~IV]{Myers02}.

This paper is organized as follows. In \S~\ref{sec:cbi} we summarize
the important properties of the CBI and introduce the mosaic
technique. In \S~\ref{sec:obs} we describe the observations that are
presented in this paper and present images of the three pairs of
mosaic fields. In \S~\ref{sec:ps} we describe the maximum-likelihood
method for estimating the power spectrum from visibility measurements
and present our power-spectrum estimates. We pay particular attention
to the contaminating effects of foreground point sources. Finally, in
\S~\ref{sec:concl} we discuss some of the implications of our results
and summarize our conclusions. A full discussion of the implications
for cosmology will be the subject of two further papers
(\citealt{Sievers02}, hereafter Paper~V; \citealt{Bond02}, hereafter
paper~VI).

\section{The Cosmic Background Imager}
\label{sec:cbi}

Theoretical models of the CMB predict its angular power spectrum,
\begin{equation}
C_\ell = \langle \vert a_{\ell m}\vert^2\rangle,
\end{equation}
where $a_{\ell m}$ are the coefficients in a spherical-harmonic
expansion of the CMB temperature distribution as a function of
direction $\vect{x}$,
\begin{equation}
{T(\vect{x}) - T_0\over T_0} = \sum_{\ell=0}^{\infty} \sum_{m=-\ell}^{\ell} a_{\ell m} Y_{\ell m}(\vect{x}),
\end{equation}
and $T_0 \approx 2.725$~K is the mean CMB temperature
\citep{Mather99}.  The angle brackets indicate the expectation value
(ensemble average).  In this paper, as in most work, the quantity
presented in the figures is the power per unit logarithmic interval in
$\ell$,
\begin{equation}
{\cal C}_{\ell} \equiv \ell(\ell+1) C_\ell / 2\pi,
\end{equation}
scaled by $T_0^2$ to put it in temperature units ($\mu$K$^2$).

Measurement of the CMB power spectrum with interferometers has been
discussed in several papers
\citep[e.g.,][]{Hobson95,Maisinger97,White99a,White99b,Ng01,Hobson02},
and details of the method that we have used are presented in Paper~IV.
A single-baseline interferometer
is sensitive to a range of multipoles $\ell \approx 2\pi u \pm
\Delta\ell/2$ where $u$ is the baseline length in wavelengths, and
$\Delta\ell$ is the FWHM of the visibility window function, which is
proportional to the square of the Fourier transform of the primary
beam (antenna power pattern).  For a circular Gaussian primary beam of
FWHM $\theta_{\rm FWHM}$ rad, $\Delta\ell = 4 \sqrt{2} \ln 2/ \theta_{\rm FWHM}$.

The CBI is a 13 element interferometer in which all 78 antenna pairs
are cross-correlated (for a detailed description, see
\citealt{Padin02}). Its 26--36~GHz band is split into ten channels
each 1~GHz wide, which are correlated separately, giving a total of
780 complex visibility measurements in each integration. The antennas
are arranged with a common axis on a flat platform. The platform mount is steered in
altitude and azimuth so that all the antennas track the same point on
the celestial sphere; in addition, the platform is rotated about the
axis to track parallactic angle, so that each baseline keeps
a constant orientation relative to the field of view. The 78 baselines
range in length from 1.0~m to about 5.5~m, depending on the antenna
configuration on the platform.  During the observations reported here,
we used several different configurations. The antennas respond to left
circular polarization (LCP), although for part of the observations one
antenna was configured for right circular polarization (RCP). Data
from the 12 cross-polarized baselines have not been used for this
paper, but they will be used to place limits on CMB polarization
(J. K. Cartwright et al., in preparation).

\begin{figure}
\plotone{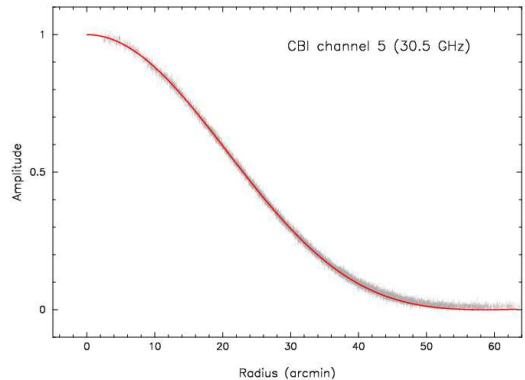}
\caption{Radial profile of the CBI primary beam in one of the ten
frequency channels; data from all 78 baselines are superimposed. The
observations ({\it error bars}) were made on 2000 Nov 12; Tau~A was
observed for about 20~s at each point on a $13\times13$ grid in
azimuth and elevation. The grid points were separated by 7\arcmin, so
the grid extended to $\pm42\arcmin$. Measurements within 45\arcmin\ of
the central position were used to fit for the amplitude scale and
pointing offset (3 parameters) for each of the 780 datasets (78
baselines $\times$ 10 channels). The {\it red curve} shows the adopted
profile, which was computed by taking the square of the Fourier
transform of the aperture illumination pattern, assumed to be
circularly symmetric.  Taking the outer radius of the aperture
(0.45~m) as $r=1$, the inner part $r < 0.172$ is blocked by the
secondary and is assumed to have zero illumination. The illumination
is tapered from center to edge: we approximate this as a Gaussian
$\exp[-(r/r_0)^2]$.  The parameter $r_0 = 0.683$ was chosen to give
the best fit to all ten channels. It corresponds to an edge taper of
$-18.6$ dB and a beam FWHM of $\theta_{\rm FWHM} = 45\farcm2 \times
(31\,{\rm GHz}/\nu)$. The 13 antennas have very similar beams,
but they have relative pointing offsets of 1--2\arcmin, up to
5\arcmin\ in the worst case.}
\label{fig:beam}
\end{figure}

The CBI Cassegrain antennas have a diameter of 0.90~m and a measured
primary beam width $\theta_{\rm FWHM} = 45\farcm2 \times (31\,{\rm
GHz}/\nu)$ at frequency $\nu$, so that $\Delta\ell \approx 300$. The
primary beam is quite close to a circular Gaussian, but we have
adopted a more accurate model of the radial profile (see
Fig.~\ref{fig:beam}) and used this model when making images and
estimating the CMB power spectrum.

In observations from a single pointing (as in Paper~I and Paper~II),
the resolution in $\ell$ of the power spectrum is limited to $\sim
\Delta\ell$, which is insufficient to resolve the expected structure
in the spectrum. To improve the resolution in $\ell$, which is
inversely proportional to the angular size of the imaged region, we
make {\it mosaiced} observations in which we map a larger area of sky
using several closely-spaced pointings. This method is in widespread
use for making images of extended regions with radio interferometers
\citep[e.g.,][]{CORNWELL88, SAULT96}. In the observations reported
here, we mapped three separate mosaic fields in this way, using 42
pointings for each in a rectangular grid of 7 rows separated by
$20\arcmin$ in declination and 6 columns separated by
$1\minute20\second \approx 20\arcmin$ in right ascension. This allows
us to improve the resolution in $\ell$ to $\Delta\ell \approx 100$.

Although the CBI antennas were designed to have low sidelobes and
crosstalk \citep{Padin00}, emission from the ground contaminates the
data, especially on short baselines (\citealt{Padin02}; Paper~I;
Paper~II). The ground signal is stable on time-scales of many minutes,
so if we observe two nearby fields under similar ground conditions,
the difference of the visibilities of the two fields is unaffected by
the ground. Differencing also eliminates any constant or slowly
varying instrumental offsets. We observe a field (the {\it lead}
field) for about 8~min and then switch to a reference field ({\it
trail} field), at the same declination but 8~min later in right
ascension, for the next 8~min, and form the difference of corresponding
$8.4$~s integrations.  One such ``scan'' consists of up to 50
differenced integrations, the exact number depending on how much time is
lost to slewing and calibration.  The two fields are observed over the
same range of azimuth and elevation, so they have nearly identical
ground contributions.  All the results presented in this paper are
derived from the differenced visibilities. The images show the
difference of intensity between a region of sky and one 8~min later in
right ascension; and the differencing is included in the covariance
matrices used for power-spectrum estimation.

\begin{figure*}
\plotone{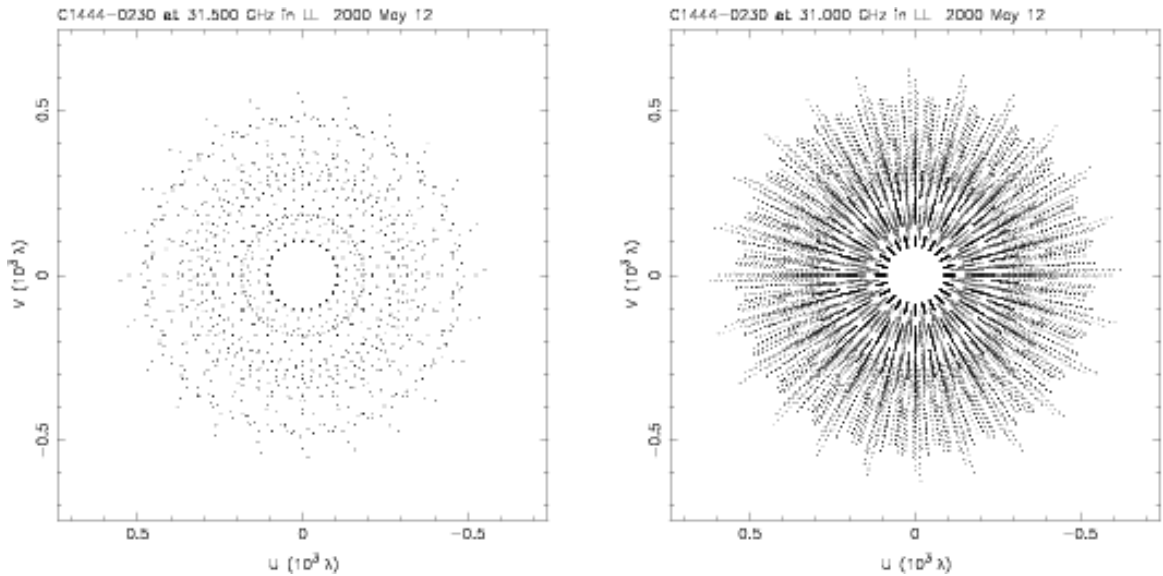}
\caption{Typical $(u,v)$ sampling obtained for a single pointing
(C1444$-$0230). For this observation only 12 antennas were used. {\it
Left}: sampling for a single frequency channel. {\it Right}: sampling
with 10 channels; a separate dot has been used for each channel. The
different channels are sensitive to slightly different angular
scales.}
\label{fig:uvplot}
\end{figure*}

For each pointing incorporated in the mosaics we obtained
approximately 16 such 8~min {\it lead-trail} scans, although the exact
number varied from one pointing to another. Between scans, we rotated
the platform to change its orientation relative to the hour circle,
thus improving the sampling of the $(u,v)$ plane\footnote{The vector
$\vect{u} \equiv (u,v)$ is the separation of a pair of antennas,
measured in wavelengths ($\lambda$), in a plane perpendicular to the
direction of the center of the field of view, i.e., in the plane of
the rotating antenna platform.}. This also reduces the effect of any
residual ground contamination: the ground signal does not add
coherently when the visibilities from different baselines measuring
the same $(u,v)$ point are combined, because the antennas have
different far sidelobe responses. An example of the $(u,v)$ plane
sampling obtained for a single pointing is shown in
Figure~\ref{fig:uvplot}. No attempt was made to obtain identical
sampling for all the pointings.

\newpage

\section{Observations}
\label{sec:obs}

\subsection{Summary of the dataset}

In this paper we present observations of three mosaic fields
(identified as 02\hour, 14\hour, and 20\hour) separated by about
6\hour\ in right ascension at a declination of about $-3\fdg5$.  The
fields were chosen to have IRAS 100~$\mu$m emission less than
1~MJy$\,$sr$^{-1}$, low synchrotron emission, and no point sources
brighter than a few hundred mJy at 1.4~GHz \citep{NVSS}. Details of
the fields observed are given in Table~\ref{tab:fields}. Two of the
individual pointings comprising our mosaics have been observed to much
greater depth, and our data for these two pointings are a subset of
the data analyzed in Paper~II.

Data acquisition, calibration, and editing were performed in the
manner described in Paper~II; we give only a summary
here. Observations were made at elevations $>42\arcdeg$, at night, and
more than $60\arcdeg$ from the moon. The amplitude scale was based on
nightly observations of calibration sources (Jupiter, Saturn, Tau~A
[3C~144, the Crab Nebula], and Vir~A [3C~274]); the primary calibrator
was Jupiter, for which we assumed an effective
temperature\footnote{This is the excess brightness over the CMB,
expressed as a temperature using the Rayleigh-Jeans approximation.} of
$152\pm5$~K at 32~GHz \citep{Mason99}. We estimate that the overall
calibration uncertainty is $5\%$ rms, equivalent to $10\%$ in CMB
power. A small fraction of the data were discarded owing to
instrumental problems. Most of these problems were detected by
real-time monitoring of the receivers; a few hardware problems in the
correlators were indicated by unusually high correlation between the
real and imaginary parts of the visibility. A few nights were affected
by bad weather, and we deleted all data taken at times when
atmospheric noise was visible on the short baselines.

As mentioned above, all the data were taken in pairs of 8~min scans on
a {\it lead} and a {\it trail} field, separated by 8 min in RA. The field separation on the sky varied with declination, but for our declinations, $\approx -3\fdg5$, 
it was very close to $2\arcdeg$. Individual 8.4~s integrations in
the two scans were matched in hour angle and differenced, and
unmatched integrations were discarded. We estimated the noise in each
scan from the rms of the differenced integrations, and compared it with the expected rms noise level in either the real or imaginary visibility:
\begin{equation}
\sigma = {\sqrt{2} k_{\rm B} T_{\rm sys} \over A \eta_Q \sqrt{\Delta\nu\tau}},
\end{equation}
where $k_{\rm B}$ is Boltzmann's constant, $T_{\rm sys}$ is the system
temperature, $A$ is the effective area of each antenna, $\eta_Q$ is
the correlator efficiency, $\Delta\nu$ is the channel bandwidth, and
$\tau$ is the integration time \citep{TMS}.  Under good conditions the
measured noise, $\sigma \approx 4.7$ Jy~s$^{1/2}$, is consistent with
the estimated system temperature $\approx 30$~K. The rms noise in the
differenced data should be $\sqrt{2}\sigma$. When the rms exceeded 2.6 times
the expected rms, we
discarded the entire scan-pair. This eliminated almost all of the data
affected by the atmosphere.  A final visibility estimate for each
channel of each baseline in each orientation of the antenna platform
was computed by a weighted average of the individual scan
visibilities, and the uncertainty in this estimate was computed from
the scan rms's, taking into account a bias introduced by the fact that
the scan rms's are themselves estimated from the data (see Paper~II).

The final edited and calibrated dataset is summarized in
Table~\ref{tab:fields}, which reports the total integration time
(averaged over baselines) on each of the lead and trail fields
comprising the mosaics.  Owing to the vagaries of the weather and the
telescope, some fields were observed to greater depths than others,
and a few were missed altogether; this is reflected in the variation
of sensitivity across each mosaic (see \S~\ref{sec:images}).

\subsection{Foreground Point Sources}

In the 26--36~GHz band, the dominant confusing foreground is the
emission from discrete radio galaxies and quasars, which we refer to
as ``point sources'' (they are virtually unresolved by the CBI). The
contribution of point sources to the visibilities must be removed
in order to obtain a reliable estimate of the CMB power
spectrum. A random distribution of point sources has a power spectrum
$C_\ell ={\rm constant}$, while for the CMB $C_\ell$ decreases rapidly
with increasing $\ell$, so discrete sources dominate  at high~$\ell$.

To remove most of the point-source contamination in our data, we
measured the flux densities of a large number of known point sources
in the mosaic fields using a new dual-beam 31~GHz HEMT receiver with a
beamwidth of $81\arcsec$ on the 40-meter telescope at the Owens Valley
Radio Observatory (OVRO). The observations followed standard methods
(see, e.g., \citealt{Myers93}); details will be presented elsewhere
(B. S. Mason et al., in preparation).  A total of 2225 sources
brighter than 6~mJy at 1.4~GHz in the NRAO VLA 1.4 GHz Sky Survey
(NVSS; \citealt{NVSS}) were observed, as described in Paper~II. For
each source detected above a $4\sigma$ threshold, the expected
visibility (assuming a point source at the NVSS position with the OVRO
flux density attenuated by the CBI primary beam) was subtracted from
the CBI visibility data. The survey is 90\% complete for $S_{31\,\rm
GHz} > 16$~mJy and 99\% complete for $S_{31\,\rm GHz} > 21$~mJy. A
total of 70 sources were subtracted from the 02\hour\ data, 63 from
the 14\hour\ data, and 68 from the 20\hour\ data. To estimate the
response in each CBI frequency channel, we used the two-point spectral
index $\alpha$ (1.4--31~GHz) for steep-spectrum sources ($\alpha <
-0.5$ where $S_\nu \propto \nu^\alpha$) and an average spectral index
$\alpha=-0.23$ for the remainder, which may be variable; but as we are
only extrapolating $\pm16\%$ in frequency the results are not very
sensitive to the choice of $\alpha$. 

By making images before and after source subtraction (see
\S~\ref{sec:images}) we have verified that the CBI and OVRO
measurements are consistent. The resulting images are dominated by the
CMB. However, source subtraction is not sufficient for accurate
power-spectrum estimation, because there may be small residuals and
there will also be unmeasured sources that have not been accounted
for.  When estimating the CMB power spectrum, we have adopted a more
powerful method than source subtraction: we have projected out the
known point sources (following \citealt{Halverson02}) (see
\S~\ref{sec:sources}). Although we subtracted the sources measured at
OVRO, we also projected them out, so errors in the OVRO flux-density
measurements should not affect the power spectrum estimates.

\subsection{Images}
\label{sec:images}

The quantity of primary cosmological interest, the power spectrum, is
best estimated directly from the visibility data, as discussed in
\S~\ref{sec:ps} below. However, we can also make images of the CMB
(convolved with the instrumental point-spread function) by
Fourier-transforming the visibilities. The images provide a good check
for the presence of non-gaussian features in the CMB or instrumental
errors in the data, such as calibration errors (which would show up as
residuals after point-source subtraction).

We have used standard aperture-synthesis techniques \citep{VLABOOK} to
make images from each pointing. The
images are formed from linear combinations of the measured
visibilities and we have not done any deconvolution or ``cleaning.''
The ``dirty'' image $I_D$ is the Fourier transform of the sampled
visibilities,
\begin{equation}
I_D(\vect{x}) = { \sum_k w_k \left\{ V_k^R \cos[2\pi\vect{u}_k\cdot\vect{x}] 
                              -V_k^I \sin[2\pi\vect{u}_k\cdot\vect{x}] \right\}
             \over \sum_k w_k }, 
\end{equation}
and is the convolution of $A(\vect{x})I(\vect{x})$ (the primary beam response
times the sky brightness) with the point-spread function or dirty beam
$B_D(\vect{x})$:
\begin{equation}
B_D(\vect{x}) = { \sum_k w_k \cos[2\pi\vect{u}_k\cdot\vect{x}]
             \over \sum_k w_k }. 
\end{equation}
In practice, we resample the visibilities on a grid and use an FFT to
compute the image and beam using standard software \citep{Difmap}.  The
sky coordinates $\vect{x}$ are the Fourier conjugates to the baseline
components $\vect{u}$ and correspond to direction cosines relative to
the pointing center. The weights $w_k$ are usually chosen to be the
statistical weights (``natural weighting''), $w_k = 1/\sigma_k^2$
where $\sigma_k$ is the standard deviation of the real or imaginary
part of the complex visibility, estimated as described above. With
this weighting, the variance of the dirty image is
\begin{equation}
s^2 = \langle{[I_D(\vect{x}) - \langle{I_D(\vect{x})}\rangle]^2}\rangle = {1\over\sum_k \sigma_k^{-2}}.
\end{equation}

A ``linear'' mosaic image $I_M(\vect{x})$ is formed from several
pointings by shifting each dirty image to a common phase center,
correcting it for its primary beam, and making a weighted sum at each
pixel:
\begin{equation}
I_M(\vect{x}) = {\sum_p W_p(\vect{x}) { I_{D,p}(\vect{x}) /  A_p(\vect{x}) } \over
            \sum_p W_p(\vect{x}) }.
\end{equation}
The weighted mean gives a maximum-likelihood estimate if the weights
are equal to the inverse variances of the corrected images, i.e.,
$W_p(\vect{x}) = A_p^2(\vect{x})/s^2_p$, where $s^2_p$ is the variance of the
dirty image made from the $p$th pointing.
Thus
\begin{equation}
I_M(\vect{x}) = {\sum_p A_p(\vect{x}) I_{D,p}(\vect{x}) /  s_p^2  \over
            \sum_p A_p^2(\vect{x}) / s_p^2 },
\label{eq:mosaic}
\end{equation}
and the variance in the mosaic image is
\begin{equation}
\sigma_M^2(\vect{x}) = {1 \over \sum_p A_p^2(\vect{x}) / s_p^2 }
\label{eq:noise}
\end{equation}
\citep{CORNWELL88, SAULT96}. The variance varies across the image, and
it is necessary to truncate the image where the variance becomes
excessive.\footnote{The full covariance matrix of the image pixels can
be calculated in a similar way, but as we estimate the power spectrum from the
visibility data directly we do not need this for our analysis.}
We have used the model for the primary beam described in
the caption to Figure~\ref{fig:beam}, set to zero for radii
$>90\arcmin$. 
In making the mosaic images, and in the power spectrum
analysis described below, we have made the approximation that the sky
is flat over the $\sim 3\arcdeg$ image, which is equivalent to an
error of $<5\arcdeg$ of phase on the longest baselines.

\begin{figure*}
\plotone{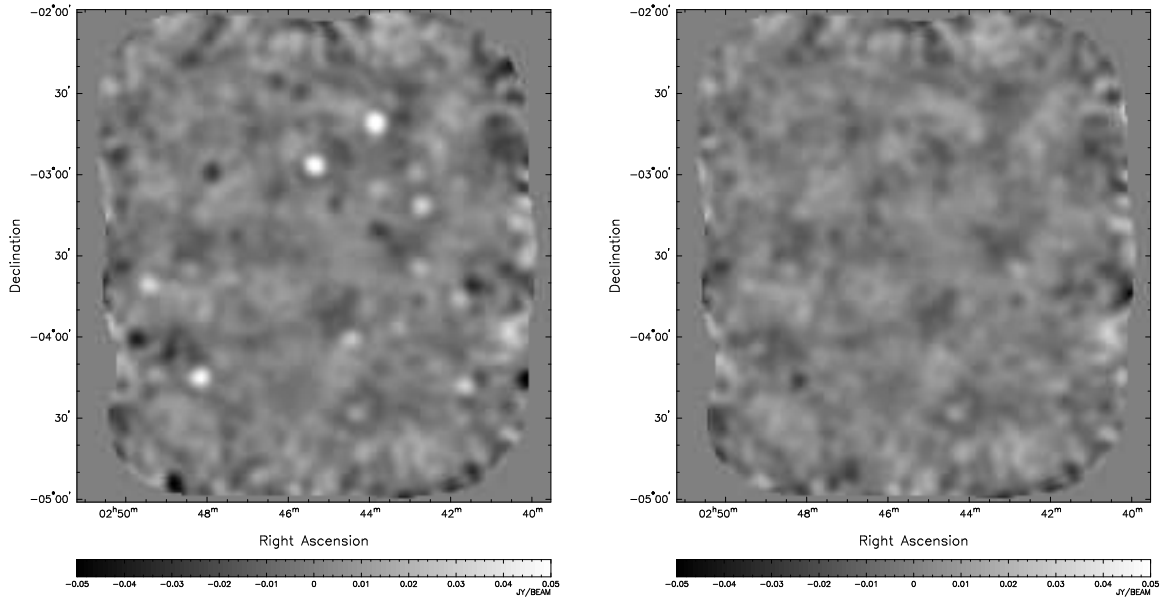}
\caption{Images of the 02\hour\ mosaic. {\it Left}: raw data;
{\it right}: after subtraction of sources measured at OVRO. The images
show the difference of the emission in the {\it lead} and {\it trail}
fields. The coordinates are J2000. The right ascension scale applies
to the {\it lead} field; add 8\minute\ to obtain the right ascension
of objects in the {\it trail} field. The same gray-scale range has
been used for both images, and it does not show the full brightness range
of the discrete sources.  Light (positive) spots are discrete sources
in the {\it lead} field, while dark (negative) spots are discrete
sources in the {\it trail} field. The brightest source in this image has a
flux density of about 67~mJy at 31~GHz. These images were made from
the entire dataset using natural weighting, and they have been corrected
for the primary beam response as described in \S~\ref{sec:images}. They
have a resolution (FWHM) of $5\farcm2$--$5\farcm5$ (the resolution
varies slightly across the image, depending on the $(u,v)$ coverage
obtained for each pointing).}
\label{fig:mos02}
\end{figure*}

\begin{figure*}
\plotone{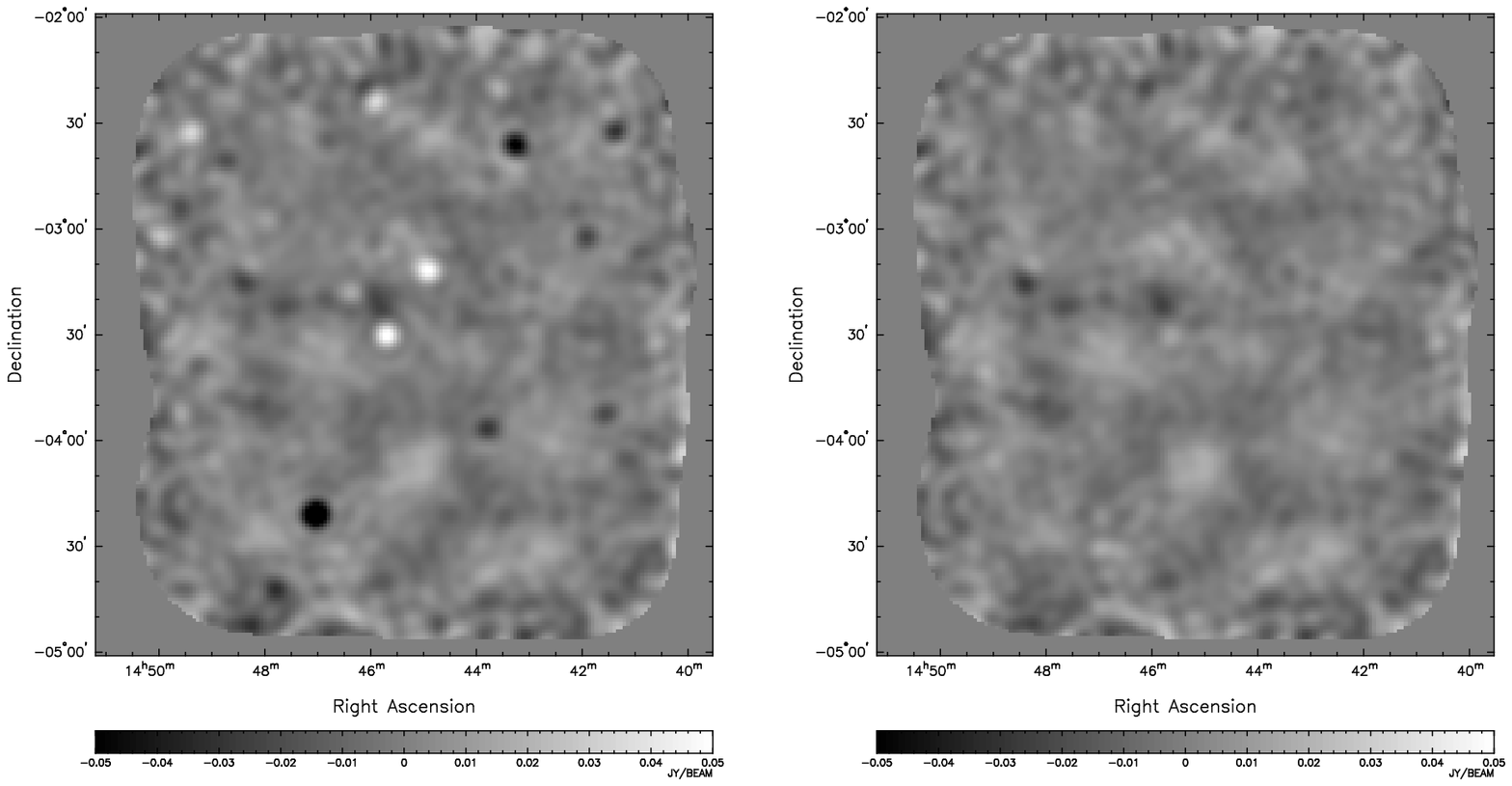}
\caption{Images of the 14\hour\ mosaic. {\it Left}: raw data;
{\it right}: after subtraction of sources measured at OVRO. For
details, see caption to Fig.~\ref{fig:mos02}.}
\label{fig:mos14}
\end{figure*}

\begin{figure*}
\plotone{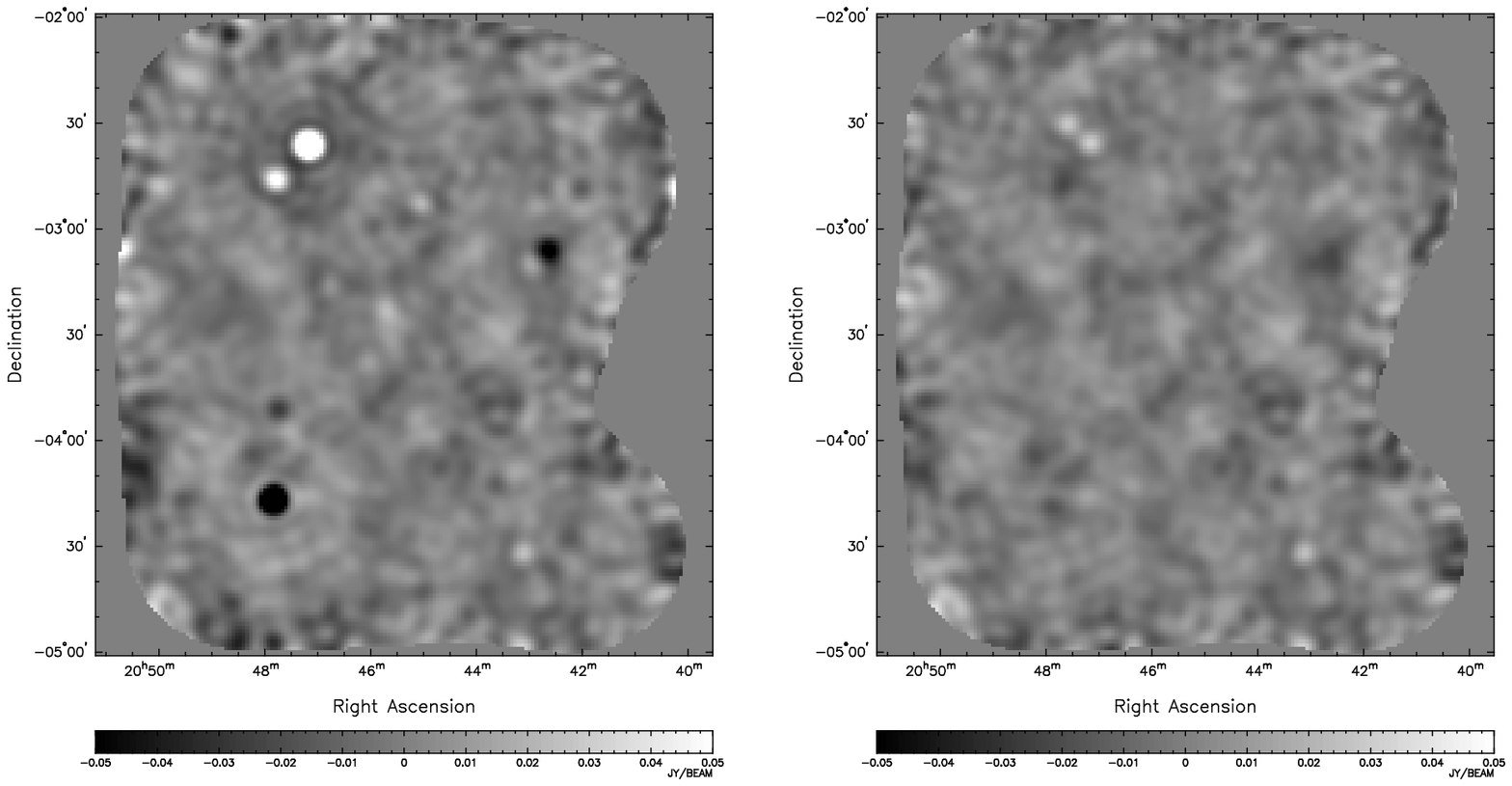}
\caption{Image of the 20\hour\ mosaic. {\it Left}: raw data; {\it right}:
after subtraction of sources measured at OVRO.  For
details, see caption to Fig.~\ref{fig:mos02}.}
\label{fig:mos20}
\end{figure*}

\begin{figure*}
\plotone{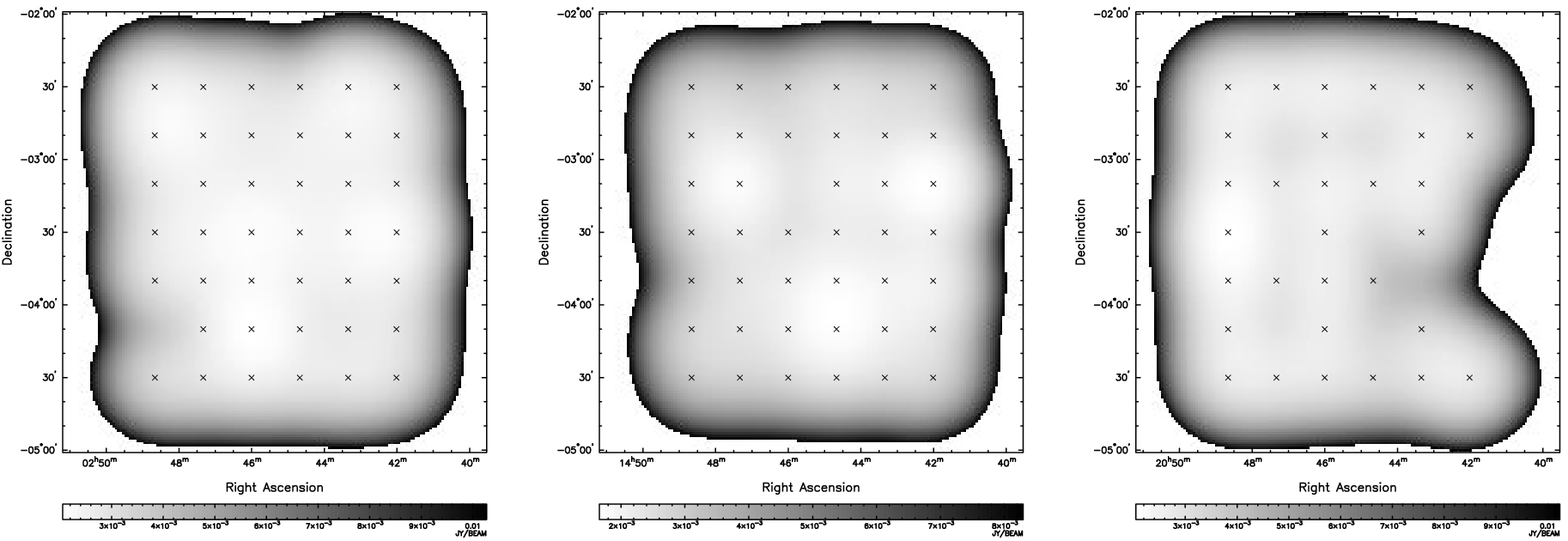}
\caption{The variation of rms noise level $\sigma_M$ across each
mosaic image (see equation~\ref{eq:noise}). The pointing centers are
marked with $\times$.  The outer parts of the images, where the noise
level exceeds a specified threshold, have been blanked. In the
02\hour\ and 20\hour\ mosaics, the minimum noise level is 2.0 mJy/beam and the
blanking threshold is 10 mJy/beam. The corresponding numbers are 1.7 and 8.3
mJy/beam for the 14\hour\ mosaic. The sky areas covered within these thresholds are 7.45 (02\hour), 7.13 (14\hour), and 7.20 (20\hour) deg$^2$ in each of {\it lead} and {\it trail}.}
\label{fig:noise}
\end{figure*}

\begin{figure*}
\plotone{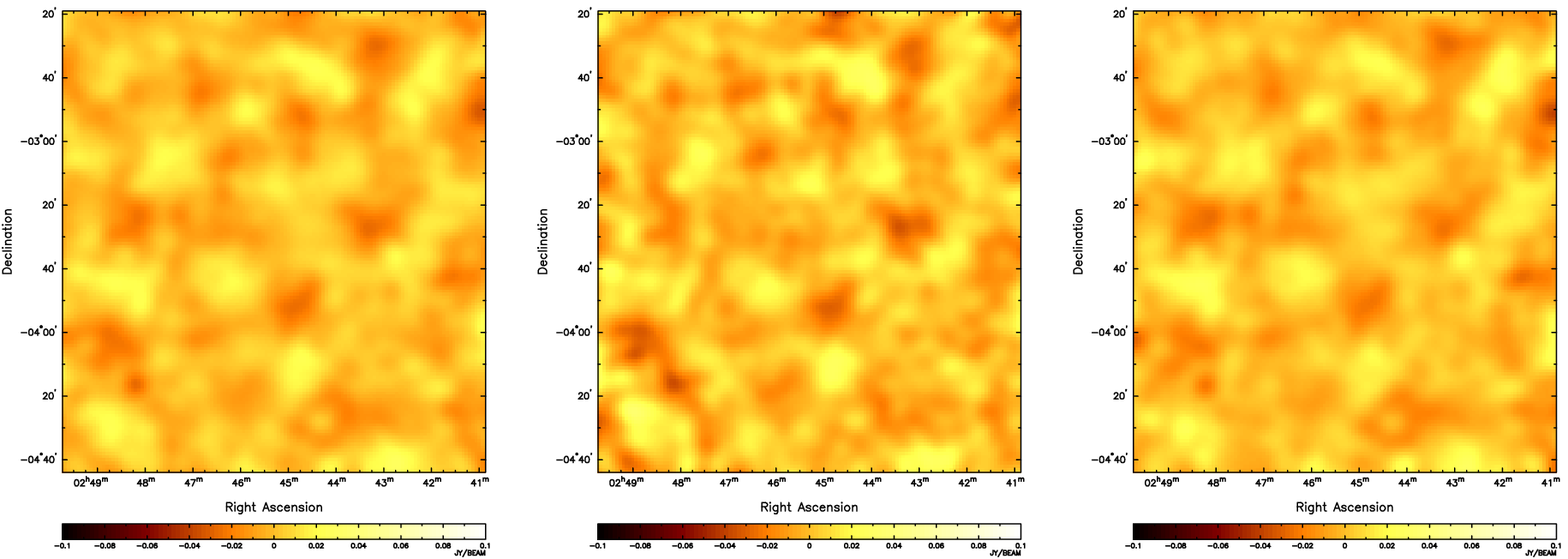}
\caption{Central part of the 02\hour\ mosaic, imaged with lower
resolution, after subtraction of sources measured at OVRO. A Gaussian
taper has been applied to the visibilities, falling to 0.1 at a
baseline length of 500 wavelengths.  {\it Left}: all frequency
channels (26--36 GHz); beam FWHM $\approx 6.9\arcmin$. {\it Center}:
high-frequency channels (31--36 GHz); beam FWHM $\approx
6.6\arcmin$. {\it Right}: low-frequency channels (26--31 GHz); beam
FWHM $\approx 7.1\arcmin$.}
\label{fig:mos02lr}
\plotone{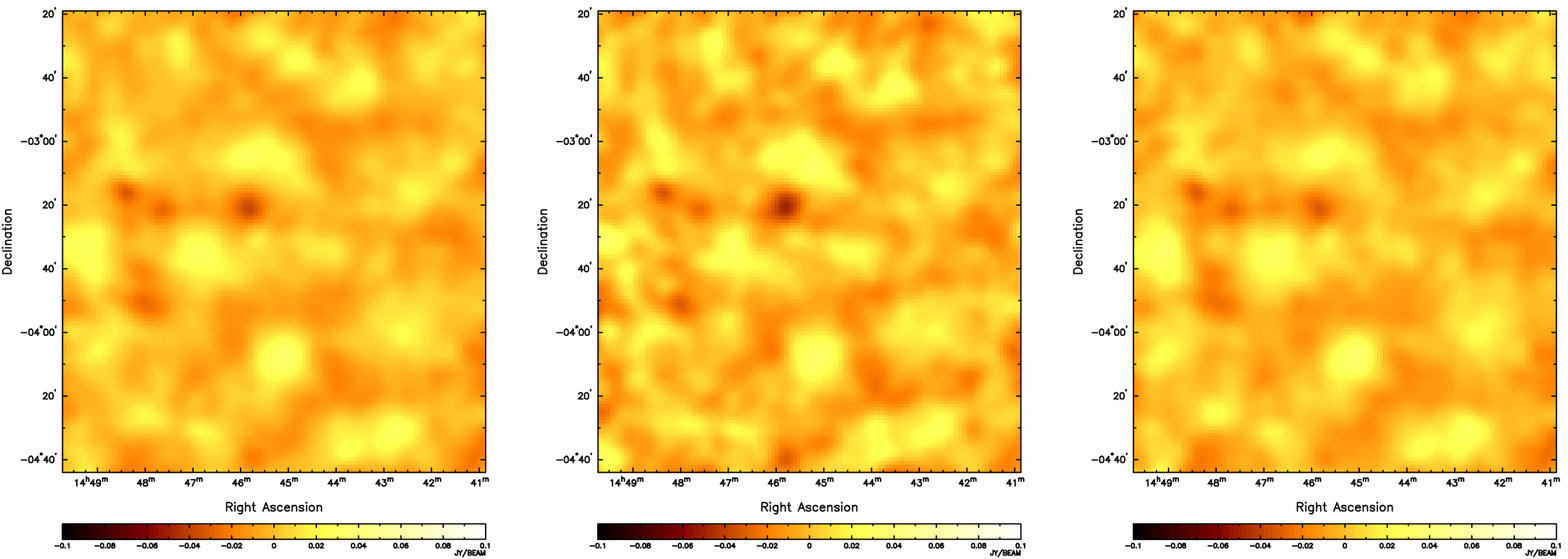}
\caption{Central part of the 14\hour\ mosaic, imaged with lower
resolution, after subtraction of sources measured at OVRO. A Gaussian
taper has been applied to the visibilities, falling to 0.1 at a
baseline length of 500 wavelengths.  {\it Left}: all frequency
channels (26--36 GHz); beam FWHM $\approx 6.9\arcmin$. {\it Center}:
high-frequency channels (31--36 GHz); beam FWHM $\approx
6.6\arcmin$. {\it Right}: low-frequency channels (26--31 GHz); beam
FWHM $\approx 7.1\arcmin$.}
\label{fig:mos14lr}
\plotone{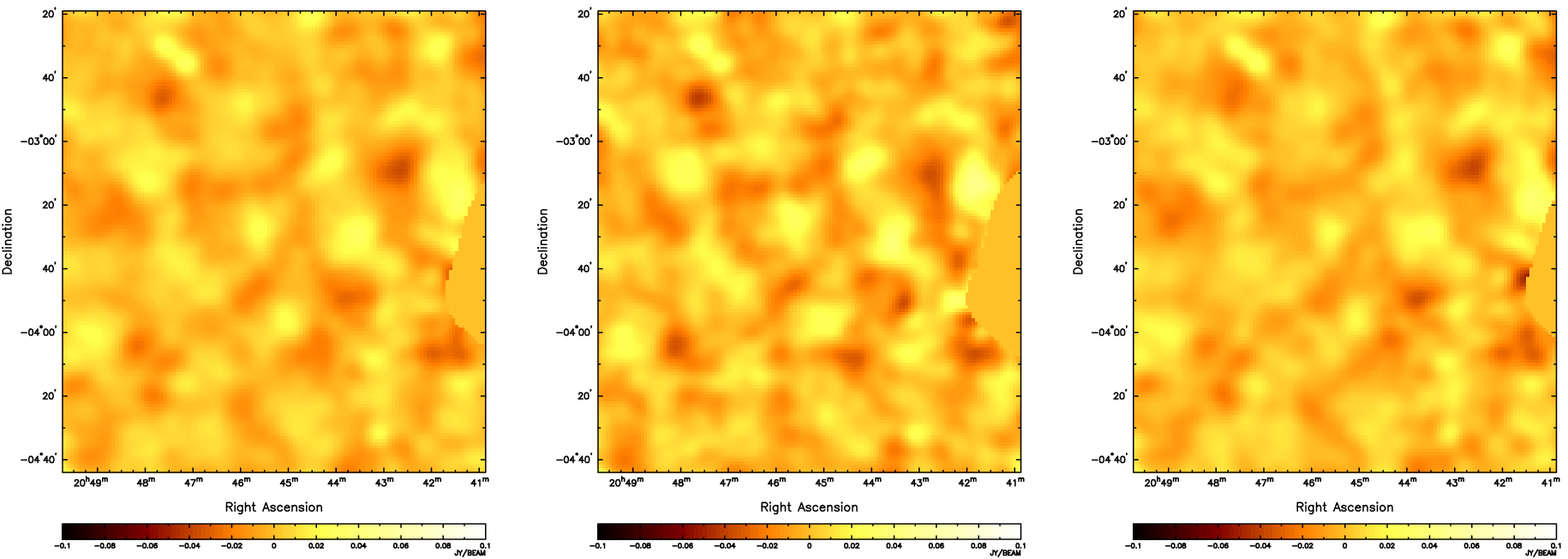}
\caption{Central part of the 20\hour\ mosaic, imaged with lower
resolution, after subtraction of sources measured at OVRO. A Gaussian
taper has been applied to the visibilities, falling to 0.1 at a
baseline length of 500 wavelengths.  {\it Left}: all frequency
channels (26--36 GHz); beam FWHM $\approx 6.9\arcmin$. {\it Center}:
high-frequency channels (31--36 GHz); beam FWHM $\approx
6.6\arcmin$. {\it Right}: low-frequency channels (26--31 GHz); beam
FWHM $\approx 7.1\arcmin$.}
\label{fig:mos20lr}
\end{figure*}

Figures \ref{fig:mos02}--\ref{fig:mos20lr} show mosaic images made
from the CBI data. We can make a variety of different images by
selecting different subsets of the data and adjusting the weights
$w_k$.

Figures \ref{fig:mos02}, \ref{fig:mos14}, and \ref{fig:mos20} show
mosaic images made from the entire dataset using natural weighting,
and similar images made from the visibility data after subtraction of
the point sources measured at OVRO. Figure \ref{fig:noise} shows the
variation of noise level across these images, and indicates the
pointing centers. The effectiveness of the OVRO source subtraction can
be seen by comparison of the ``before'' and ``after'' images. Most of
the sources have been removed successfully, although there are of
course residuals owing to measurement errors or source
variability, and a few sources can be seen that were not detected at
OVRO (these sources were projected out in the power spectrum
analysis).

To investigate the sources further, we have made higher-resolution
images (not shown) using only baselines longer than 250$\lambda$, and
searched for peaks exceeding $5\sigma_M$ in images of signal-to-noise
ratio. All these peaks are coincident within $2\arcmin$ with sources
that have  $S _{1.4\,\rm GHz}> 3.4$~mJy in the NVSS catalog.  Of the 42
sources detected, 37 are identified with NVSS sources with
$S_{1.4\,\rm GHz} > 6$~mJy which were detected at OVRO.  The remaining
five objects are associated with NVSS sources that were not detected
at OVRO; one of them, with $S_{1.4\,\rm GHz} < 6$~mJy, was not observed
at OVRO.  As we reported in Paper~II, we find that the number $N$ of
sources greater than flux density $S$ at 31~GHz is
\begin{equation}
N(S) \approx 2.8 \pm 0.7 \,{\rm deg}^{-2} \left( S\over 10\,{\rm mJy}\right)^{-1.0}
\end{equation}
for $5 < S < 50$~mJy. We use this result in \S~\ref{sec:sources} to
estimate the contribution to our power spectra of sources below
$S_{1.4\,\rm GHz}> 3.4$~mJy (stronger sources are treated individually).

Figures \ref{fig:mos02lr}, \ref{fig:mos14lr}, and \ref{fig:mos20lr}
show lower-resolution images made by reducing the weight of long
baselines with a Gaussian taper; this suppresses the high-frequency
noise and emphasizes the CMB emission. These figures include images
made from the upper and lower halves of the CBI frequency band. There
is good visual correspondence between the two halves of the band; but
note that the spatial-frequency sampling is not the same in the two
halves, so a quantitative comparison is difficult. The visual
agreement is echoed in the good agreement between the power spectra
obtained from the two halves of the band (see
\S~\ref{sec:sub_by_freq}). These low-resolution images are dominated
by sky signal rather than noise. Note, however, that the images are
differences of two sky patches, and they are also missing the lowest
spatial frequencies, so a comparison with images obtained with another
instrument may be difficult.  By inspection of signal-to-noise ratio
images (computed using Eqs.~\ref{eq:mosaic} and \ref{eq:noise}) made
from subsets of the data, we have found that signals corresponding to
$\ell \lesssim 1000$ are detected with high significance ($\gtrsim
3\sigma$), and there are some significant detections of individual
features at higher $\ell$.  The detected features range in angular
size from $\sim 6\arcmin$ to $\sim 15\arcmin$, corresponding to mass
scales at the surface of last scattering of $\sim 5 \times 10^{14}$ to
$8 \times 10^{15}\,M_\sun$, so these features are likely to be the
seeds that would evolve into clusters of galaxies by the present
epoch.

\newpage

\section{Power Spectrum}
\label{sec:ps}

\subsection{Algorithm}

Our algorithm for the estimation of power spectra from mosaic
visibility data is described in Paper~IV. We model the power
spectrum as flat in each of a set of $N_B$ contiguous bands, i.e.,
\begin{equation}
 {\cal C}_\ell \equiv \ell(\ell+1)C_\ell/2\pi = q_B, \quad \ell_{B-1} < \ell < \ell_B
\label{eq:bins}
\end{equation}
(with $\ell_0 = 0$), and take the {\it band-powers} $q_B$
($B=1,\ldots,N_B$) as a set of unknown parameters to be determined by
maximizing the likelihood (the probability of obtaining the measured
visibilities, if the model were correct, for given values of the
parameters).  If the signal and noise obey Gaussian statistics, which
we assume, the likelihood is given by the multivariate
Gaussian distribution for  {\it complex} variates,
\begin{equation}
{\cal L}(q_B) = {1\over \pi^n \det \tjptensor{C}} \exp\left(- \vect{V}^\dag \tjptensor{C}^{-1} \vect{V} \right),
\label{eq:like}
\end{equation}
where $\vect{V}$ is a column vector containing the complex visibility
measurements, $\tjptensor{C} = \langle \vect{V} \vect{V}^\dag\rangle$ (a
function of the parameters $q_B$) is the covariance matrix of the
visibilities, and ${}^\dag$ denotes the Hermitian conjugate. (Although
eq.~[\ref{eq:like}] is expressed in terms of complex visibilities, it
is easier in practice to treat the real and imaginary parts of the
visibilities as a double-length real vector).

The correlation matrix $\tjptensor{C}$ can be written as
\begin{eqnarray}
\tjptensor{C} & = & \tjptensor{C}^{\rm N} 
                      + \sum_B q_B \tjptensor{C}_B^{\rm S}
                      + q_{\rm res} \tjptensor{C}^{\rm res}  \nonumber\\  
              & &         + q_{\rm OVRO} \tjptensor{C}^{\rm OVRO}
                          + q_{\rm NVSS} \tjptensor{C}^{\rm NVSS},
\label{eq:cvm}
\end{eqnarray}
where $\tjptensor{C}^{\rm N}$ is the noise correlation matrix, estimated
from the data as described above, $\tjptensor{C}_B^{\rm S}$ is the CMB
signal correlation matrix for band $B$ (independent of $q_B$),
$\tjptensor{C}^{\rm OVRO}$ and $\tjptensor{C}^{\rm NVSS}$ are constraint
matrices \citep{BJK98} representing the effects of foreground point
sources of known position, and $\tjptensor{C}^{\rm res}$ represents a
residual contribution from faint sources of unknown position. We
discuss the source terms further in \S~\ref{sec:sources}. The factors
$q_{\rm res}$, $q_{\rm OVRO}$, and $q_{\rm NVSS}$ could in principle
be regarded as free parameters to be determined by maximum likelihood,
but in practice they are not well determined by the data and we
instead hold them fixed at {\it a priori} values.

In a typical mosaic observation, the number of distinct visibility
measurements is very large ($\sim 200\,000$ complex visibilities for a
mosaic of 42 pointings with 10 frequency channels, 78 baselines, and
several parallactic angles), which makes the covariance matrices
(eq.~[\ref{eq:cvm}]) impractically large. However, neighboring points
in the $(u,v)$ plane are highly correlated and need not be treated
completely independently.  To reduce the size of the matrices, we
interpolate the measured visibility set $\vect{V}$ onto a smaller
number of grid points in the $(u,v)$ plane (so that the quantities that
enter the likelihood calculation are linear combinations of the
measured visibilities), and make the corresponding transformation of
the covariance matrix.  By this means we reduce the
dimension of the (real) matrices to $\lesssim 5000$. We 
ran tests with different grid spacings, using both real data and 
simulated data, to verify that using a finer grid would not 
significantly change the results (see Paper~IV). We find the
maximum likelihood solution by the quadratic relaxation technique of
\citet{BJK98}, which yields estimates of the band powers $q_B$, with
their covariances given by the inverse of the Fisher information
matrix
\begin{equation}
F_{BB'} = \left\langle {\partial^2 \ln {\cal L} \over \partial q_B \partial q_{B'}}\right\rangle = {1\over 2} {\rm Tr}\left(\tjptensor{C}^{-1} \tjptensor{C}_B^{\rm S}\tjptensor{C}^{-1} \tjptensor{C}_{B'}^{\rm S} \right).
\end{equation}
We also calculate the band-power window functions and the equivalent
band-powers of the noise, known point sources, and residual point
sources by the methods described in Paper~IV. The band-power window
function $W^B(\ell)$ \citep{Knox99} allows the expected band-power for
a given model power spectrum to be estimated as a weighted mean:
\begin{equation}
\langle q_B \rangle = \sum_\ell {\cal C}_\ell {W^B(\ell)\over \ell}.
\label{eq:winfunc}
\end{equation}

\subsection{Correction for Foreground Point Sources}
\label{sec:sources}

If we had accurate measurements of all the point sources in
our frequency band, we could subtract their contributions directly
from the visibilities, but as we do not we must adopt a statistical
approach.  A detailed description of our method of dealing with point
sources is presented in Paper~II, and here we give only a
summary.

The strongest sources (``OVRO'' sources) were measured at OVRO and
subtracted from the visibility data as described above. The
subtraction was necessarily imperfect, however, and the residuals of
the subtracted sources make a contribution to the power spectrum that
must be removed.  We created a second list of ``NVSS'' sources within
about $60\arcmin$ of any of the mosaic pointing centers that were not
detected at OVRO but had $S_{1.4\,\rm GHz} > 3.4$~mJy, which
corresponds to $S_{31\,\rm GHz} > 0.25$~mJy for a typical spectral
index of $-0.84$.  This list contained 960, 918, and 974 sources in
the 02\hour, 14\hour, and 20\hour\ mosaic fields. Crude estimates of
the flux densities for these sources were subtracted from the
visibilities as part of the power-spectrum estimation. For each
mosaic, we constructed two constraint matrices, $\tjptensor{C}^{\rm
OVRO}$ and $\tjptensor{C}^{\rm NVSS}$, using the positions and estimated
residual flux-density uncertainties of the sources in the two lists.
In principle, if our error estimates were correct, the constraint
matrices would fully account for the source contributions and should
be included in the maximum likelihood analysis with prefactors $q_{\rm
OVRO} = q_{\rm NVSS} = 1$. However, after some experimentation, we
decided to err on the side of caution and use large pre\-factors,
$q_{\rm OVRO} = q_{\rm NVSS} = 10^5$, to give very low weight to modes
that are affected by these known sources (much larger factors cause
the covariance matrix to be ill-conditioned). This is equivalent to
marginalizing over the unknown flux densities, or ``projecting out''
the sources \citep{Halverson02, BJK98}. Our analysis is thus
insensitive to errors in the assumed flux densities of the sources,
but at the cost of some loss in sensitivity (see Paper IV).

Sources that are not included in the known-source lists also
contribute power. We estimate the visibility covariance arising from
such sources and include it as the residual source term
$\tjptensor{C}^{\rm res}$. The radio source counts and spectra used to
compute this term are described in Paper~II. As in that paper, we
estimate that, for an NVSS flux-density cutoff at $S_{1.4\,\rm GHz} >
3.4$~mJy, the amplitude of the residual correction is $C_\nu^{\rm res}
= 0.08 \pm 0.04$~Jy$^2$~sr$^{-1}$ (see Paper~IV).  The dividing line
between known sources and sources included in the residual term is
somewhat arbitrary, so long as the residual term is computed correctly
for the chosen flux-density cutoff. We have conducted tests to verify
that our results are insensitive to the precise choice of cutoff. For
a full discussion of the source projection and the residual
correction, see Paper~II.

\begin{figure}
\plotone{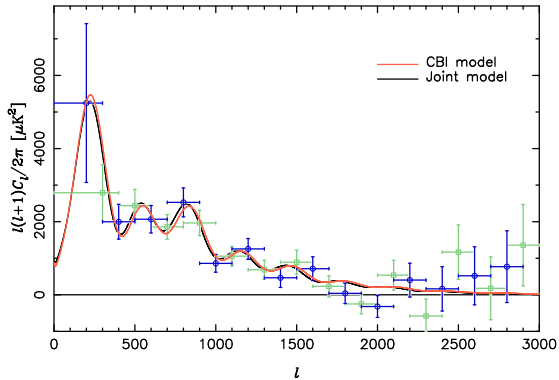}
\caption{Joint power spectrum estimates for the three CBI mosaics.
Band-power estimates have been made for two alternate
divisions of the $\ell$ range into bins: ``even'' binning ({\it green
squares}) and ``odd'' binning ({\it blue circles}). The error-bars
show $\pm1\sigma$ uncertainties from the inverse Fisher matrix.  Two
minimal inflation-based models are shown.
{\it Red:} fit to CBI plus {\it COBE} DMR;
$\Omega_{\rm tot}=1.0$, $\Omega_{b}h^2 = 0.0225$, $\Omega_{\rm cdm}h^2
= 0.12$, $\Omega_\Lambda= 0.6$, $n_s=0.95$, $\tau_c=0.025$, ${\cal
C}_{10} = 786$ $\mu$K$^2$.
{\it Black:} joint fit to CBI, DMR, DASI, BOOMERANG-98, VSA and earlier data;
$\Omega_{\rm tot}=1.0$, $\Omega_{b}h^2 = 0.02$,
$\Omega_{\rm cdm}h^2=0.14$, $\Omega_\Lambda= 0.5$, $n_s=0.925$,
$\tau_c=0$, ${\cal C}_{10} = 887$ $\mu$K$^2$. For details,
see Paper~V.}
\label{fig:jointps200}
\end{figure}

\begin{deluxetable}{lrc}
\tabletypesize{\scriptsize} 
\tablecolumns{3} 
\tablewidth{0pc} 
\tablecaption{Band-Powers and Uncertainties\label{tab:bandpowers}} 
\tablehead{\colhead{$\ell$ range}&
\colhead{$\ell_{\rm eff}$}&
\colhead{\begin{tabular}{c}
Band-Power $l(l+1)C_l/(2\pi)$\\
($\mu$K$^2$)
\end{tabular}}}
\startdata 
0--400&      304&2790  $\pm$  771\\
400--600&    496&2437  $\pm$  449\\
600--800&    696&1857  $\pm$  336\\
800--1000&   896&1965  $\pm$  348\\
1000--1200& 1100&1056  $\pm$  266\\
1200--1400& 1300& 685  $\pm$  259\\
1400--1600& 1502& 893  $\pm$  330\\
1600--1800& 1702& 231  $\pm$  288\\
1800--2000& 1899&$-$250$\pm$  270\\
2000--2200& 2099& 538  $\pm$  406\\
2200--2400& 2296&$-$578$\pm$  463\\
2400--2600& 2497&1168  $\pm$  747\\
2600--2800& 2697& 178  $\pm$  860\\
2800--3000& 2899&1357  $\pm$ 1113\\
\tableline
0--300&      200&5243  $\pm$ 2171\\
300--500&    407&1998  $\pm$  475\\
500--700&    605&2067  $\pm$  375\\
700--900&    801&2528  $\pm$  396\\
900--1100&  1002& 861  $\pm$  242\\
1100--1300& 1197&1256  $\pm$  284\\
1300--1500& 1395& 467  $\pm$  265\\
1500--1700& 1597& 714  $\pm$  324\\
1700--1900& 1797&  40  $\pm$  278\\
1900--2100& 1997&$-$319$\pm$  298\\
2100--2300& 2201& 402  $\pm$  462\\
2300--2500& 2401& 163 $\pm$  606\\
2500--2700& 2600& 520  $\pm$  794\\
2700--2900& 2800& 770  $\pm$  980\\
\enddata
\end{deluxetable}

\subsection{Results}

\subsubsection{Joint Mosaic Power Spectrum}
\label{sec:jointmosaic}

\begin{figure}
\plotone{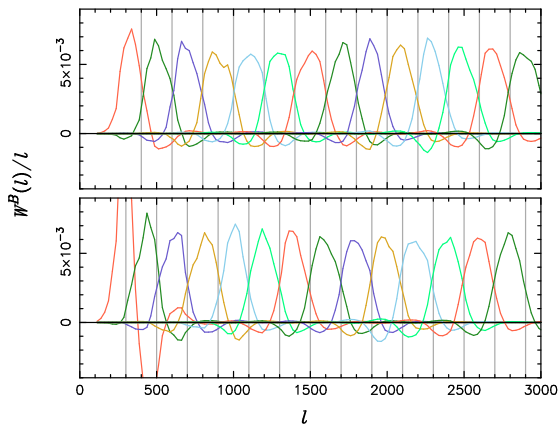}
\caption{Window functions for the joint power spectrum estimates for
the three CBI mosaics. {\it Top}: ``even'' binning. {\it Bottom}:
``odd'' binning. The vertical grey lines show the bin boundaries. Each function
is normalized: $\sum_\ell W^B(\ell)/\ell =1$.}
\label{fig:jointwf200}
\end{figure}

\begin{figure}
\plotone{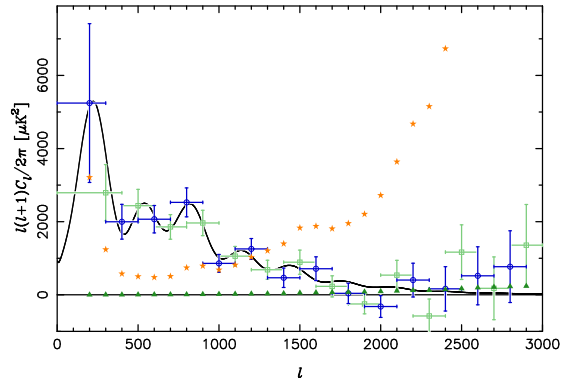}
\caption{Joint power spectrum estimates for the three CBI mosaics
 (the same as Fig.~\ref{fig:jointps200}). The {\it stars}
represent the power spectrum of the instrumental noise correction, and
the {\it triangles} represent the power spectrum of the residual
source correction. The {\it black curve} is the joint model from
Figure~\ref{fig:jointps200}.}
\label{fig:jointcomp}
\end{figure}

The primary result of this paper is the power spectrum of the CMB in
the three mosaics treated jointly. For this analysis we have
estimated the power spectrum in bins of width $\Delta \ell =200$, with
two alternate locations of the bins. The ``even'' binning has $\ell_B
= 200 + 200 B$ ($1 \le B \le 16$), while the ``odd'' binning has
$\ell_B = 100 + 200 B$ ($1 \le B \le 16$); here $\ell_B$ is the upper
limit of the bin, as in equation~(\ref{eq:bins}). In both cases the
first bin is wider and starts at $\ell=0$. While we included bins at
higher $\ell$, we report the results only for $\ell < 3000$: at higher
$\ell$, the mosaic data have very little sensitivity.  The two sets of
bins are of course not independent. The results are given in
Table~\ref{tab:bandpowers}, which gives for each bin the band-power
$q_B$, the rms uncertainty in $q_B$ from the Fisher matrix, and the
centroid of the window function $\ell_{\rm eff}$.  The results are
also displayed in Figure~\ref{fig:jointps200}, and the window
functions are shown in Figure~\ref{fig:jointwf200}.\footnote{The window
functions and inverse Fisher matrices are available on the CBI web
page, \url{http://www.astro.caltech.edu/~tjp/CBI/}.}  With
$\Delta\ell = 200$, the adjacent bins are anticorrelated by about
$16\%$. We have also computed power-spectrum estimates using narrower
bins with $\Delta\ell=140$, for which the anticorrelation of adjacent bins is
about $24\%$,
again using overlapping ``odd'' and ``even'' bins. We have used all
four binnings ($\Delta\ell=140$ odd and even, and $\Delta\ell=200$ odd
and even) for cosmological-parameter estimation (see Paper~V), and all
four give consistent results.  The component band-powers (defined in
Paper~IV) for instrumental thermal noise ($\tjptensor{C}^{\rm N}$) and the residual
source correction ($\tjptensor{C}^{\rm res}$) are shown in
Figure~\ref{fig:jointcomp}.  This figure shows that the residual
source correction is negligible at $\ell \lesssim 2000$; the thermal
noise, however, exceeds the signal for $\ell > 1300$ and is the
dominant effect at high $\ell$.  Both these corrections increase
approximately, but not exactly, as $\ell^2$.  The thermal noise
depends on the sampling in the $(u,v)$ plane and better-sampled $\ell$
bins have lower noise; while the residual source correction has a
non-thermal spectrum and the magnitude of its contribution in any bin
depends on the spectral sensitivity in that bin---this varies from bin
to bin as the CBI frequency channels do not have the same $(u,v)$
sampling.

Figure~\ref{fig:jointps200} shows two theoretical spectra for minimal
inflation-based models with different parameters. As we discuss in
Paper~V, we have evaluated the posterior probabilities of the CBI and
other datasets over a grid of models in a 7-dimensional parameter
space, using a variety of prior probabilities based on Hubble
constant, large-scale-structure, and supernova-Ia observations. The
first model displayed in Figure~\ref{fig:jointps200} is the model that
maximizes the posterior probability of the CBI mosaic 
(using the ``odd'' $\Delta\ell=140$ binning) and {\it COBE}-DMR
results, with the weak-$h$ prior on the Hubble constant; within the grid
of models, this is also the best fit if we restrict the search to flat
models.  The second model is intended to represent a current
``concordance'' model: it is the best fit of the CBI, DMR, DASI, and
BOOMERANG-98 data with flat and weak-$h$ priors. We find the same
model if we also include LSS, SN, or HST-$h$ priors. The parameters
for the two models are given in the figure caption. It is remarkable
that the two spectra are so similar. The CBI data together with DMR
place strong constraints on the allowed region of parameter space. These
constraints are consistent with those obtained from earlier CMB
observations, even though the CBI is sampling an $\ell$ range a factor
of two larger than that spanned by the earlier experiments.  For a
full discussion of parameter estimation from the CBI results, see
Paper~V.

\begin{figure*}
\epsscale{0.80}
\plotone{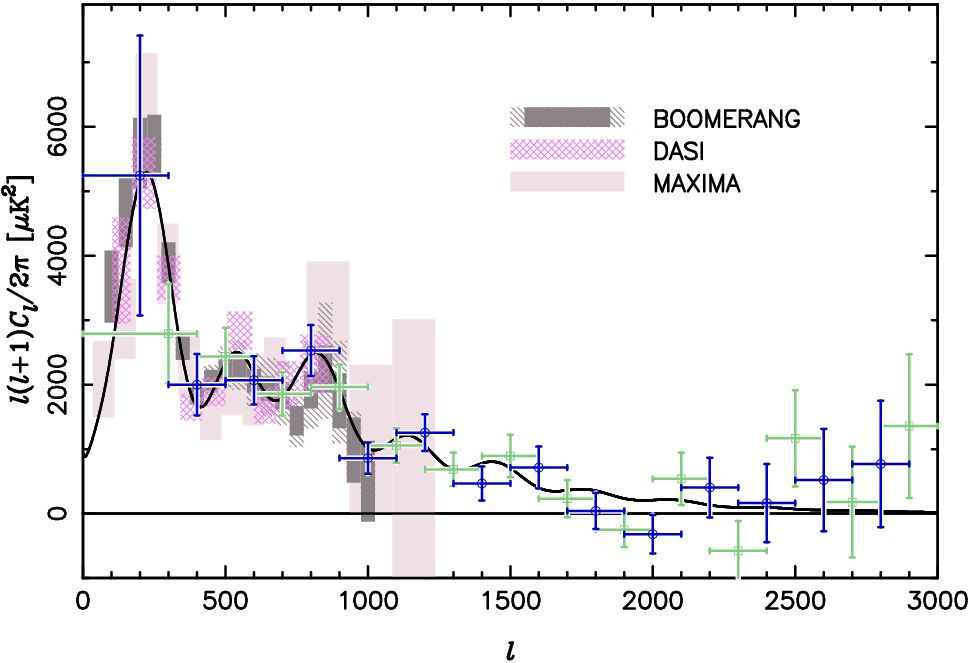}
\caption{Comparison of the joint power spectrum estimates from the
 three CBI mosaics (Fig.~\ref{fig:jointps200}) with the measurements
 from BOOMERANG \citep{Netterfield02}, DASI \citep{Halverson02}, and
 MAXIMA \citep{Lee01}. The rectangles indicate the 68\% confidence
 intervals on band-power; for BOOMERANG, the solid rectangles indicate
 the 68\% confidence interval for the statistical and sample variance
 errors, while the hatched rectangles shows the amount by which a
 $\pm1\sigma$ error in the beamwidth ($12\farcm9 \pm 1\farcm4$) would
 shift the estimates (all up or all down together). The {\it black
 curve} is the joint model from Figure~\ref{fig:jointps200}.}
\label{fig:compare}
\end{figure*}

We compare our results with the earlier results from the BOOMERANG,
DASI, and MAXIMA experiments in Figure~\ref{fig:compare}. In the
region of overlap ($300\lesssim \ell \lesssim 1000$) the agreement is
very good. A detailed comparison will be made in Paper~V.

\begin{figure}
\plotone{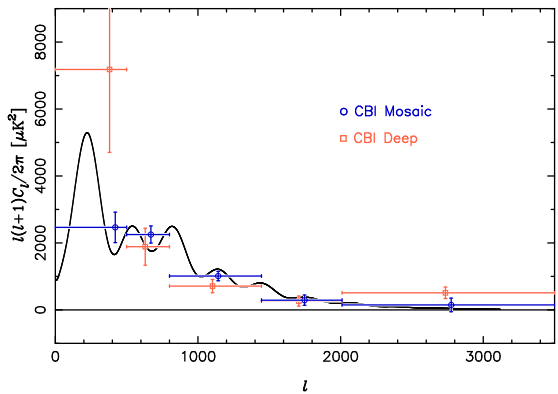}
\caption{Comparison of the joint power spectrum estimates from the
 three CBI mosaics ({\it blue circles}) with those from the
 three deep fields of Paper~II ({\it red squares}); both have been
 computed for the $\ell$ bins used in Paper~II. 
 The {\it black curve} is the joint model from
Figure~\ref{fig:jointps200}.}
\label{fig:mos_deep}
\end{figure}

We have also computed the power spectrum of the three mosaics
using the wider bins chosen for analyzing the CBI deep field data
(Paper~II). The deep and mosaic results are compared in
Figure~\ref{fig:mos_deep}.  We have used six bins: $\ell = 0$--500,
500--800, 880--1445, 1445--2010, 2010--3500, and $\ell > 3500$. The
last bin contains virtually no data for the mosaics, and we have
omitted it from the figure.  The first bin ($\ell < 500)$) is poorly
constrained by the data and has a large uncertainty owing to sample
variance.  Two of the three deep fields lie within the area covered by
the mosaics, so the results are not entirely independent. But if we
ignore this complication, we can compare the two sets of band-powers
by a chi-squared test, using the band-power covariance matrices. For
this test, we assume that the likelihood function is approximately
Gaussian and compute
\begin{equation}
\chi^2 = \sum_B \sum_{B'}
         \left( q_B^{(1)} -q_B^{(2)} \right)
         \left( M_{BB'}^{(1)} + M_{BB'}^{(2)} \right)^{-1}
         \left( q_{B'}^{(1)} -q_{B'}^{(2)} \right),
\label{eq:chisq}
\end{equation}
where $q_B^{(1)}, q_B^{(2)}$ are the band-power estimates and
$M_{BB'}^{(1)}, M_{BB'}^{(2)}$ are the inverse Fisher matrices for the
two datasets.  Omitting the first and last bins, we find
$\chi^2=5.57$, with 5 degrees of freedom. If the two data sets were
drawn from the same population, a larger value would be obtained in
35\% of trials, so we conclude that the two data sets are consistent.
At $\ell > 2000$, where the deep observations show a 
significant signal, the mosaic observations are less sensitive than
the deep and are consistent both with the deep result and with no
signal. In the bin $2010 < \ell < 3500$, we find $q_B = 148\pm203$
$\mu$K$^2$ in the mosaics and $q_B = 510\pm171$ $\mu$K$^2$ in the deep
fields. The thermal noise band-power in this bin is much larger for
the mosaic data set than for the deep data set, so the deep results
are less sensitive to systematic errors in the noise estimation and
are thus more reliable than the mosaic results.

\subsubsection{Peaks and Dips in the Power Spectrum}

The power spectra with bin size $\Delta\ell=200$
(Fig.~\ref{fig:jointps200}) suggest the presence of peaks and dips, but,
owing to the anticorrelations between adjacent bins, their reality is
difficult to assess in this presentation, or in similar plots of the
$\Delta\ell=140$ bins. To assess their significance, we have searched
for extrema in the power spectrum following 
techniques applied to the BOOMERANG data by \cite{debernardis02}. For
each triplet of adjacent bins $[i-1,i,i+1]$ we model the local band-power
profile as a three-parameter quadratic form
\begin{equation}
q_B = a {\ell}^2_B - 2 b {\ell}_B + c , \quad\quad B=i-1,i,i+1, 
\end{equation}
where ${\ell}^2_B$, ${\ell}_B$ are band-average values of $\ell^2$ and
$\ell$.  In terms of the fitted parameters $a,b,c$, the peak location
is $\ell_{\rm pk} = b/a$, its amplitude is ${\cal C}_{\rm pk} = c -b^2/a$, and
its curvature $\kappa_{\rm pk} = a$. We have assumed that the measured
$q_B$'s are Gaussian-distributed with a covariance
$F_{BB^\prime}^{-1}$. In this Gaussian approximation for the
likelihood ${\cal L}(q_B$), the likelihood ${\cal L}(a,b,c)$ of the
quadratic parameters is also a Gaussian: the maximum values
$a_m,b_m,c_m$ are a direct transform of the data band-power averages in
the three bins, and the curvature at the maximum, which describes the
uncertainty in these parameter estimates, is simply related to
$F_{BB^\prime}$.  The maximum-likelihood values of $\ell_{\rm pk}$, ${\cal
C}_{\rm pk}$ and $\kappa_{\rm pk}$ are determined by $a_m,b_m,c_m$, but the errors
in these transformed variables are non-Gaussian. We estimate the errors
by computing the local curvature of the likelihood near maximum, by a
Jacobian transformation.  We consider a peak or a dip in the spectrum
to be detected if two conditions are met: (1) the position $\ell_{\rm pk}$
is within the range of multipoles covered by the band triplet; (2) the
best fit quadratic has a curvature $a_m$ which differs from zero by at
least $1\sigma$.

\begin{figure}
\plotone{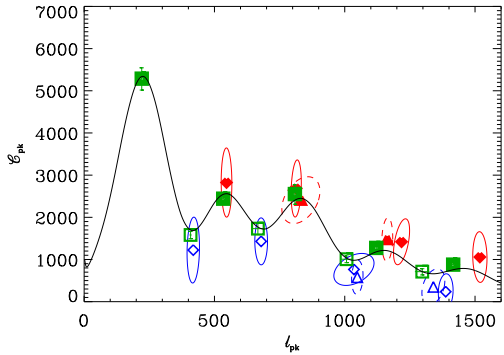}
\caption{Results for the peak and dip detection algorithm applied to
the $\Delta\ell=140$ data. Peaks ({\it red}) and dips ({\it blue})
found in the odd binning are indicated by diamonds with ellipses
showing the $1\sigma$ confidence contours; those found in the even
binning are shown by triangles with dashed confidence contours.  Only
detections with curvature above $1\sigma$ are shown.
For comparison, the {\it filled} and {\it open green squares} show the
peak and dip positions and amplitudes predicted from earlier CMB
observations in the context of inflation-motivated models: they are
from ensemble averages over the ${\cal C}_\ell$-database we use for
parameter estimation in Paper~V (for the weak-$h$ prior).
In this figure the {\it black curve} is the CBI+DMR model from
Figure~\ref{fig:jointps200}.}
\label{fig:peaksdips}
\end{figure}

The results of applying this algorithm to the $\Delta \ell=140$ odd
and even binned data for $\ell < 2000$ are shown in
Figure~\ref{fig:peaksdips}.  There are more detections in the odd
binning than in the even binning, which is an indication that even the
$\Delta\ell=140$ bin size is larger than we would like for effective
peak/bin detection.  In the odd binning, we detect four peaks, at
$\ell \sim 550, 800, 1150, 1500$, and four dips, at $\ell \sim
400,700,1050,1400 $.  The significance levels of the peaks, expressed
as $|a_m|/\sigma$, are 2.1, 2.6, 1.8, and 1.7, and those of the dips
are 2.3, 2.4, 1.5, and 2.3.  The two peaks and two dips detected with
the even binning show good agreement with those detected with the odd
binning.  Maximum-likelihood values for the subthreshold even-binned
detections (with curvature $< 1\sigma$) are in good accord with the
odd-binned detections.  Thus we have tentative detections of the
second, third, fourth, and fifth acoustic peaks. With this dataset,
even $\Delta\ell=140$ is barely fine enough to resolve the peaks, but
we should be able to do better when we include the second season of
CBI data.

A second approach to peak/dip detection was also used by
\cite{debernardis02}: given a class of theoretical models with a
sequence of peaks and dips, the statistical distribution of positions
and amplitudes can be predicted by ensemble-averaging over the full
probability, the multidimensional likelihood. We have used the same ${\cal
C}_\ell$-database as \citet{debernardis02}, which is also
the one we have used for cosmological parameter estimation in
Paper~V.  Figure~\ref{fig:peaksdips} shows the peaks and dips we
``predict'' from BOOMERANG, DASI, MAXIMA, {\it COBE} DMR and 19 other
experiments predating this CBI dataset. The errors on the positions
and heights determined this way are relatively small, comparable to
the size of the symbols plotted. Within this set of minimal
inflation-based models, the positions and amplitudes of the
higher-$\ell$ peaks are largely determined by the positions and
amplitudes of the first few. It can be seen that the values found
using our model-independent quadratic peak/dip-finder are in excellent
agreement with the predictions for $\ell < 1000$. At higher $\ell$,
perhaps there is a shift of peak placement, but we caution that our
peak-position error bars, being derived from a Fisher matrix
determined at the maximum likelihood, are only approximate.  Adding
CBI to the rest of the pre-CBI experiments gives peak positions and
amplitudes in good accord with those shown here, and indeed so does
using just DMR and the CBI data.

\begin{figure}
\plotone{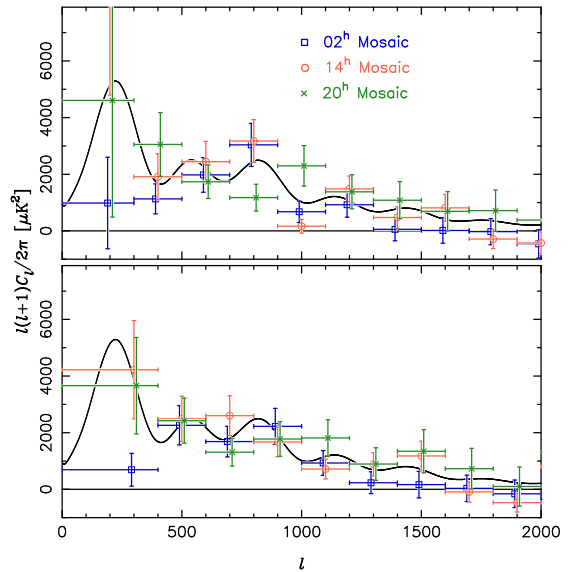}
\caption{Power spectrum estimates for the three CBI mosaics,
treated independently. The {\it upper panel} shows the results for the
odd binning, and the {\it lower panel} shows the results for the even
binning. In both binnings, the lowest bin is poorly constrained by the
data. In this bin, in particular, the sample variance is large. The
sample variance contribution to the error estimate is a fraction of
the fitted band-power, and is thus unrealistically low for the
02\hour\ mosaic which has low fitted band-power.}
\label{fig:indivps}
\end{figure}

\begin{deluxetable}{llcc}
\tabletypesize{\scriptsize} 
\tablecolumns{4} 
\tablewidth{0pt} 
\tablecaption{Chi-Squared Comparison of the Mosaic Power Spectra\label{tab:chisq}} 
\tablehead{\colhead{Mosaics}&
\colhead{Binning\tablenotemark{a}}&
\colhead{$\chi^2$ (d.o.f.)\tablenotemark{b}}&
\colhead{\begin{tabular}{c}Significance Level\\ (per cent)\tablenotemark{c}\\\end{tabular}}}
\startdata 
02\hour--14\hour& even&  7.8 (9)& 56\\
02\hour--14\hour& odd&   5.6 (9)& 78\\
02\hour--20\hour& even& 10.3 (9)& 33\\
02\hour--20\hour& odd&  17.4 (9)&  4\\
14\hour--20\hour& even&  7.6 (9)& 57\\
14\hour--20\hour& odd&  17.5 (9)&  4\\
\enddata
\tablenotetext{a}{The two alternate binnings are not independent. See \S~\ref{sec:jointmosaic}. The first bin has been omitted. The $\ell$ range is 400--2200
in the even binning and 300--2100 in the odd binning.}
\tablenotetext{b}{$\chi^2$ computed using the inverse Fisher matrix, assuming 
Gaussian likelihood.}
\tablenotetext{c}{Probability of exceeding the observed $\chi^2$ if the two data sets are drawn from the same population.}
\end{deluxetable} 

\subsubsection{Individual Mosaic Power Spectra}

To check the consistency of the three mosaic data sets, and to look
for variations in the power spectrum with direction on the sky, we
have computed the power spectra of the three mosaics
separately. The results are shown for the two alternate binnings in
Figure~\ref{fig:indivps}; the very noisy points for $\ell > 2000$ have
been omitted.  The chi-squared test (eq.~[\ref{eq:chisq}])
 shows that the 02\hour\ and 14\hour\ mosaics are consistent
with each other, but the 20\hour\ mosaic is discrepant at the 95\%
confidence level in the odd binning (see Table~\ref{tab:chisq}).
Figure~\ref{fig:indivps} shows that most of the discrepancy occurs in
two adjacent bins at $700 < \ell < 1100$. It is unlikely that this
reflects a real change in the CMB spectrum with direction, or that
it could arise from, for example, foreground contamination in the
20\hour\ mosaic, so we provisionally attribute the discrepancy to a
chance statistical fluctuation.  The three mosaics have galactic
latitudes $-53\arcdeg$ (02\hour), $48\arcdeg$ (14\hour), and
$-27\arcdeg$ (20\hour). The good agreement between the three mosaics
suggests that the CMB power spectrum is not heavily contaminated by
latitude-dependent diffuse emission from the Galaxy.  We will be able
to examine the question of field-to-field consistency more closely
when we have analyzed data from the 2001 observing season, which
extend the sky coverage by a factor of two.

\begin{figure}
\plotone{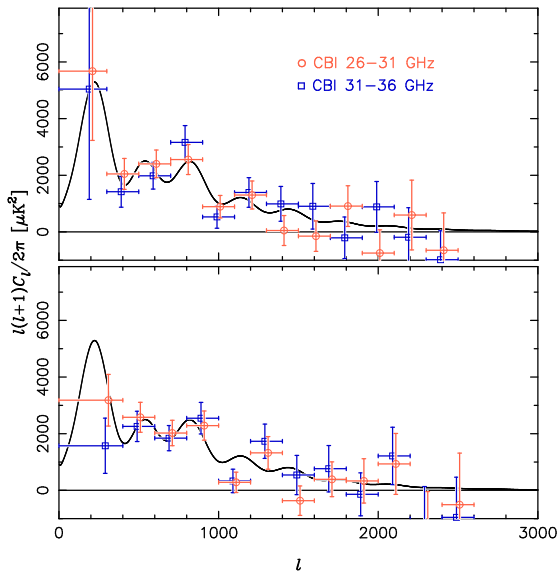}
\caption{Joint power spectrum estimates for the three CBI mosaics,
 split into the upper and lower halves of the CBI frequency
band.  The {\it upper panel} shows the results for the odd binning,
and the {\it lower panel} shows the results for the even binning.}
\label{fig:joint_bychan}
\end{figure}

\subsubsection{Subdivision by frequency}
\label{sec:sub_by_freq}

A second consistency check is to compare the power spectra obtained
from the different CBI frequency channels. As in \S~\ref{sec:images},
we have divided the data into low- and high-frequency halves (26--31
and 31--36 GHz).  The results are shown in
Figure~\ref{fig:joint_bychan}.  If the signal is due primarily to the
CMB, the two spectra should be consistent, but if there is a large
contribution from a non-thermal foreground, such as synchrotron,
free-free, or dust emission, they should be different. The two spectra
are similar, but it is difficult to make a quantitative comparison
owing to the strong correlation between the two frequency bands.  The
error estimates obtained from the Fisher matrix include the
contribution of cosmic variance as well as the measurement noise, so
they overestimate the band-to-band variance.

\section{Conclusions}
\label{sec:concl}

The results presented in this paper demonstrate the effectiveness of
interferometric mosaicing both for imaging the CMB and for measuring
its power spectrum. The CBI images show for the first time structures
in the CMB on mass-scales corresponding to clusters of galaxies, and
the CMB power spectrum has been extended by more than a factor of two
in multipole number $\ell$. Although the resolution in $\ell$ is
limited (we will obtain better resolution when we analyze the
wider-field observations made in 2001), we have been able to detect
the second and third acoustic peaks in the spectrum, and, for the
first time, the fourth and possibly the fifth.  The deeper
observations reported in Paper~II extend the spectrum even further,
well into the damping tail region where secondary anisotropies become
important (see Paper VI).

Ground-based observations of the CMB in the 1-cm wave-band, where long
integrations can be obtained, are competitive with balloon-based
observations at shorter wavelengths. The principal obstacle to
observations in this band, particularly at high $\ell$, is the
emission from foreground point sources. We have shown that it is
possible to correct the observations for this contamination with high
accuracy, at a cost in sensitivity, but foreground sources remain the
largest source of uncertainty in the power spectrum. To improve on our
result, we will need sensitive high-resolution surveys of the
foreground sources at 31 GHz, such as will soon be possible with the
NRAO Green Bank Telescope.

It should be clear from Figures~\ref{fig:jointps200} and
\ref{fig:compare} that the CBI results are consistent with earlier
observations in the region $\ell < 1000$. What is more remarkable is
that at higher $\ell$, a region that has not been probed before, the
results are consistent with extrapolations of the power spectrum based
on simple inflation-motivated models. We show in Paper V that the
major cosmological parameters (the total density parameter, the
density parameters for baryonic and non-baryonic matter, the
primordial density perturbation spectral index, the Hubble constant,
the cosmological constant, and the optical depth to last scattering)
are well constrained by the CBI observations, even when only the region
of the spectrum beyond the first two peaks is considered. This
provides further strong support for cosmological models dominated by
cold dark matter and dark energy, and with a scale-invariant spectrum
of primordial density fluctuations up to $\ell \gtrsim 2000$.  The
corresponding angular scales and masses are $\sim 6'$ and $\sim 5
\times 10^{14} M_\sun$, the scale of galaxy clusters. This provides a
firm foundation for theories of galaxy formation.

A second season of CBI mosaic observations were obtained in 2001, and
are currently being analyzed. These observations double the size of
each of the three mosaic fields, and will enable the power spectrum to
be determined with improved sensitivity and $\ell$-resolution. We are
currently reconfiguring the instrument to maximize its sensitivity to
polarization, with the goal of detecting and measuring the power
spectrum of the polarized component of the CMB, which is another
powerful test of the cosmological models.

\acknowledgements This work was supported by the National Science
Foundation under grants AST 94-13935, AST 98-02989, and AST
00-98734. Research in Canada was supported by NSERC and the Canadian
Institute for Advanced Research. The computing facilities at CITA were
funded by the Canada Foundation for Innovation.  We gratefully
acknowledge the generous support of Maxine and Ronald Linde, Cecil and
Sally Drinkward, Barbara and Stanley Rawn, Jr., and Fred Kavli, and
the strong support of the Provost and President of the California
Institute of Technology, the PMA Division Chairman, the Director of
the Owens Valley Radio Observatory, and our colleagues in the PMA
Division.  LB and JM acknowledge support from FONDECYT Grant 1010431. SC
acknowledges support by CONICYT postdoctoral grant 3010037. We thank
CONICYT for permission to operate within the Chajnantor Scientific
Preserve in Chile, Angel Otarola for invaluable assistance in setting
up the CBI in Chile, Russ Keeney for construction and operation of
the 31-GHz receiver on the OVRO 40~m telescope, and Sterl Phinney for
useful discussions.

\newpage

\LongTables

\begin{deluxetable*}{llllc} 
\tabletypesize{\scriptsize}
\tablecolumns{4} 
\tablewidth{0pt} 
\tablecaption{Summary of Observations\label{tab:fields}} 
\tablehead{\colhead{Field name\tablenotemark{a}}&
\colhead{\makebox[140pt][c]{Date}}&
\colhead{\begin{tabular}{c}R.A.\\(h, m, s)\end{tabular}}&
\colhead{\begin{tabular}{c}Decl.\\(\arcdeg, \arcmin)\end{tabular}}&
\colhead{\begin{tabular}{c}Integration\\ time (s)\\\end{tabular}}}
\startdata 
C0242$-$0230& 2000 Jul 16, Oct 20& 02\,42\,00& $-$02\,30& 4560\\
C0242$-$0250& 2000 Oct 01&         02\,42\,00& $-$02\,50& 7454\\
C0242$-$0310& 2000 Jul 31, Oct 21& 02\,42\,00& $-$03\,10& 4394\\
C0242$-$0330& 2000 Oct 04, Oct 06& 02\,42\,00& $-$03\,30& 14388\\
C0242$-$0350& 2000 Aug 03, Oct 21& 02\,42\,00& $-$03\,50& 5090\\
C0242$-$0410& 2000 Oct 18&         02\,42\,00& $-$04\,10& 6170\\
C0242$-$0430& 2000 Aug 06, Oct 22& 02\,42\,00& $-$04\,30& 4954\\

C0243$-$0230& 2000 Sep 09, Oct 22, Oct 25, Oct 26& 02\,43\,20& $-$02\,30& 10220\\
C0243$-$0250& 2000 Aug 09, Oct 22& 02\,43\,20& $-$02\,50& 5048\\
C0243$-$0310& 2000 Sep 22& 02\,43\,20& $-$03\,10& 6314\\
C0243$-$0330& 2000 Aug 29& 02\,43\,20& $-$03\,30& 5562\\
C0243$-$0350& 2000 Sep 25& 02\,43\,20& $-$03\,50& 5562\\
C0243$-$0410& 2000 Sep 02& 02\,43\,20& $-$04\,10& 3206\\
C0243$-$0430& 2000 Sep 28& 02\,43\,20& $-$04\,30& 7462\\

C0244$-$0230& 2000 Oct 20& 02\,44\,40& $-$02\,30& 2052\\
C0244$-$0250& 2000 Oct 02& 02\,44\,40& $-$02\,50& 7576\\
C0244$-$0310& 2000 Aug 01, Oct 21& 02\,44\,40& $-$03\,10& 4980\\
C0244$-$0330& 2000 Oct 07& 02\,44\,40& $-$03\,30& 6990\\
C0244$-$0350& 2000 Aug 04, Oct 21& 02\,44\,40& $-$03\,50& 5100\\
C0244$-$0410& 2000 Oct 19& 02\,44\,40& $-$04\,10& 3706\\
C0244$-$0430& 2000 Aug 07, Oct 22& 02\,44\,40& $-$04\,30& 5052\\

C0246$-$0230& 2000 Sep 10& 02\,46\,00& $-$02\,30& 6812\\
C0246$-$0250& 2000 Aug 11, Aug 12& 02\,46\,00& $-$02\,50& 3480\\
C0246$-$0310& 2000 Sep 23& 02\,46\,00& $-$03\,10& 6314\\
C0246$-$0330& 2000 Aug 30, Oct 24& 02\,46\,00& $-$03\,30& 9384\\
C0246$-$0350& 2000 Sep 26& 02\,46\,00& $-$03\,50& 6908\\
C0246$-$0410& 2000 Sep 07, Oct 25, Oct 26& 02\,46\,00& $-$04\,10& 13380\\
C0246$-$0430& 2000 Sep 29& 02\,46\,00& $-$04\,30& 6766\\

C0247$-$0230& 2000 Jul 29, Oct 20& 02\,47\,20& $-$02\,30& 4854\\
C0247$-$0250& 2000 Oct 03& 02\,47\,20& $-$02\,50& 7574\\
C0247$-$0310& 2000 Aug 02, Oct 21& 02\,47\,20& $-$03\,10& 5060\\
C0247$-$0330& 2000 Oct 08& 02\,47\,20& $-$03\,30& 7380\\
C0247$-$0350& 2000 Aug 05, Oct 22& 02\,47\,20& $-$03\,50& 4910\\
C0247$-$0410& 2000 Oct 09& 02\,47\,20& $-$04\,10& 2066\\
C0247$-$0430& 2000 Aug 08, Oct 22& 02\,47\,20& $-$04\,30& 5048\\

C0248$-$0230& 2000 Sep 11, Oct 26& 02\,48\,40& $-$02\,30& 8356\\
C0248$-$0250& 2000 Aug 15, Oct 23& 02\,48\,40& $-$02\,50& 9296\\
C0248$-$0310& 2000 Sep 24& 02\,48\,40& $-$03\,10& 4404\\
C0248$-$0330& 2000 Sep 01& 02\,48\,40& $-$03\,30& 5134\\
C0248$-$0350& 2000 Sep 27, Oct 25& 02\,48\,40& $-$03\,50& 5826\\
C0248$-$0410& \nodata    & 02\,48\,40& $-$04\,10& \nodata\\
C0248$-$0430& 2000 Sep 30& 02\,48\,40& $-$04\,30& 7170\\

C1442$-$0230& 2000 May 05 & 14\,42\,00& $-$02\,30& 7288\\
C1442$-$0250& 2000 Jul 19, Aug 23 & 14\,42\,00& $-$02\,50& 4054\\
C1442$-$0310& 2000 Apr 04, Apr 05, Apr 27 & 14\,42\,00& $-$03\,10& 26048\\
C1442$-$0330& 2000 Jul 26, Aug 17& 14\,42\,00& $-$03\,30& 4560\\
C1442$-$0350\tablenotemark{b}& 2000 Mar 17, Apr 28 & 14\,42\,00& $-$03\,50& 14610\\
C1442$-$0410& 2000 Aug 01, Aug 12, Aug 16& 14\,42\,00& $-$04\,10& 6766\\
C1442$-$0430& 2000 May 01 & 14\,42\,00& $-$04\,30& 7492\\

C1443$-$0230& 2000 Jun 24, Aug 19 & 14\,43\,20& $-$02\,30& 6152\\
C1443$-$0250& 2000 May 23 & 14\,43\,20& $-$02\,50& 5180\\
C1443$-$0310& 2000 Jun 27 & 14\,43\,20& $-$03\,10& 6814\\
C1443$-$0330& 2000 May 26 & 14\,43\,20& $-$03\,30& 5210\\
C1443$-$0350& 2000 Jul 02 & 14\,43\,20& $-$03\,50& 6402\\
C1443$-$0410& 2000 May 29 & 14\,43\,20& $-$04\,10& 7352\\
C1443$-$0430& 2000 Jul 15 & 14\,43\,20& $-$04\,30& 5582\\

C1444$-$0230& 2000 May 11 & 14\,44\,40& $-$02\,30& 8396\\
C1444$-$0250& 2000 Jul 20, Aug 18 & 14\,44\,40& $-$02\,50& 5878\\
C1444$-$0310& 2000 Apr 9, Apr 10, Apr 26 & 14\,44\,40& $-$03\,10& 13920\\
C1444$-$0330& 2000 Jul 27, Aug 22 & 14\,44\,40& $-$03\,30& 5134\\
C1444$-$0350& 2000 Apr 07, Apr 08, Apr 25 & 14\,44\,40& $-$03\,50& 13704\\
C1444$-$0410& 2000 Jun 20, Jun 21, Jun 22 & 14\,44\,40& $-$04\,10& 20112\\
C1444$-$0430& 2000 May 02 & 14\,44\,40& $-$04\,30& 6326\\

C1446$-$0230& 2000 Jun 25 & 14\,46\,00& $-$02\,30& 5066\\
C1446$-$0250& 2000 May 25 & 14\,46\,00& $-$02\,50& 5266\\
C1446$-$0310& \nodata & 14\,46\,00& $-$03\,10& \nodata\\
C1446$-$0330& 2000 May 28 & 14\,46\,00& $-$03\,30& 4904\\
C1446$-$0350& 2000 Jul 03 & 14\,46\,00& $-$03\,50& 6808\\
C1446$-$0410& 2000 May 30 & 14\,46\,00& $-$04\,10& 7366\\
C1446$-$0430& 2000 Jul 16 & 14\,46\,00& $-$04\,30& 4826\\

C1447$-$0230& 2000 May 22 & 14\,47\,20& $-$02\,30& 7254\\
C1447$-$0250& 2000 Jul 24, Jul 25 & 14\,47\,20& $-$02\,50& 4002\\
C1447$-$0310& 2000 Apr 11, Apr 12, Apr 29& 14\,47\,20& $-$03\,10& 22004\\
C1447$-$0330& 2000 Jul 29 & 14\,47\,20& $-$03\,30& 4754\\
C1447$-$0350& 2000 Apr 30, Jul 23 & 14\,47\,20& $-$03\,50& 8212\\
C1447$-$0410& 2000 Aug 15, Aug 21 & 14\,47\,20& $-$04\,10& 5166\\
C1447$-$0430& 2000 May 03 & 14\,47\,20& $-$04\,30& 4606\\

C1448$-$0230& 2000 Jun 26& 14\,48\,40& $-$02\,30& 7088\\
C1448$-$0250& 2000 May 31 & 14\,48\,40& $-$02\,50& 8490\\
C1448$-$0310& 2000 Jul 01, Aug 20 & 14\,48\,40& $-$03\,10& 5048\\
C1448$-$0330& 2000 Jun 04 & 14\,48\,40& $-$03\,30& 6308\\
C1448$-$0350& 2000 Jul 04 & 14\,48\,40& $-$03\,50& 818\\
C1448$-$0410& 2000 Jun 07 & 14\,48\,40& $-$04\,10& 8058\\
C1448$-$0430& 2000 Jul 17 & 14\,48\,40& $-$04\,30& 5222\\

C2042$-$0230& 2000 Jun 04 & 20\,42\,00& $-$02\,30& 2844\\
C2042$-$0250& 2000 Jul 28 & 20\,42\,00& $-$02\,50& 5556\\
C2042$-$0310& \nodata     & 20\,42\,00& $-$03\,10& \nodata\\
C2042$-$0330& \nodata     & 20\,42\,00& $-$03\,30& \nodata\\
C2042$-$0350& \nodata     & 20\,42\,00& $-$03\,50& \nodata\\
C2042$-$0410& \nodata     & 20\,42\,00& $-$04\,10& \nodata\\
C2042$-$0430& 2000 May 05, Jul 02 & 20\,42\,00& $-$04\,30& 9392\\

C2043$-$0230& 2000 May 31 & 20\,43\,20& $-$02\,30& 5930\\
C2043$-$0250& 2000 Jun 08 & 20\,43\,20& $-$02\,50& 7520\\
C2043$-$0310& 2000 Jul 05 & 20\,43\,20& $-$03\,10& 8024\\
C2043$-$0330& 2000 Jun 12 & 20\,43\,20& $-$03\,30& 6992\\
C2043$-$0350& \nodata     & 20\,43\,20& $-$03\,50& \nodata\\
C2043$-$0410& 2000 Jun 26 & 20\,43\,20& $-$04\,10& 7128\\
C2043$-$0430& 2000 Jul 25 & 20\,43\,20& $-$04\,30& 4356\\

C2044$-$0230& 2000 May 30 & 20\,44\,40& $-$02\,30& 7536\\
C2044$-$0250& \nodata     & 20\,44\,40& $-$02\,50& \nodata\\
C2044$-$0310& 2000 May 01 & 20\,44\,40& $-$03\,10& 4536\\
C2044$-$0330& \nodata     & 20\,44\,40& $-$03\,30& \nodata\\
C2044$-$0350& 2000 May 02 & 20\,44\,40& $-$03\,50& 2150\\
C2044$-$0410& \nodata     & 20\,44\,40& $-$04\,10& \nodata\\
C2044$-$0430& 2000 May 11 & 20\,44\,40& $-$04\,30& 4538\\

C2046$-$0230& 2000 Jul 03 & 20\,46\,00& $-$02\,30& 8724\\
C2046$-$0250& 2000 Jun 10 & 20\,46\,00& $-$02\,50& 6298\\
C2046$-$0310& 2000 Jul 06 & 20\,46\,00& $-$03\,10& 8722\\
C2046$-$0330& 2000 Jun 13 & 20\,46\,00& $-$03\,30& 7684\\
C2046$-$0350& 2000 Jul 23 & 20\,46\,00& $-$03\,50& 8302\\
C2046$-$0410& 2000 Jun 27 & 20\,46\,00& $-$04\,10& 8564\\
C2046$-$0430& 2000 Jul 26 & 20\,46\,00& $-$04\,30& 8260\\

C2047$-$0230& 2000 Jun 07 & 20\,47\,20& $-$02\,30& 7596\\
C2047$-$0250& \nodata     & 20\,47\,20& $-$02\,50& \nodata\\
C2047$-$0310& 2000 May 04 & 20\,47\,20& $-$03\,10& 3258\\
C2047$-$0330& \nodata     & 20\,47\,20& $-$03\,30& \nodata\\
C2047$-$0350& 2000 May 03 & 20\,47\,20& $-$03\,50& 2764\\
C2047$-$0410& \nodata     & 20\,47\,20& $-$04\,10& \nodata\\
C2047$-$0430& 2000 May 29 & 20\,47\,20& $-$04\,30& 7132\\

C2048$-$0230& 2000 Jul 04 & 20\,48\,40& $-$02\,30& 8312\\
C2048$-$0250& 2000 Jun 11 & 20\,48\,40& $-$02\,50& 7326\\
C2048$-$0310& 2000 Jul 07 & 20\,48\,40& $-$03\,10& 8652\\
C2048$-$0330\tablenotemark{b}& 2000 Aug 02, Aug 23& 20\,48\,40& $-$03\,30& 15316\\
C2048$-$0350& 2000 Jul 24 & 20\,48\,40& $-$03\,50& 8318\\
C2048$-$0410& 2000 Jul 01 & 20\,48\,40& $-$04\,10& 6660\\
C2048$-$0430& 2000 Jul 27 & 20\,48\,40& $-$04\,30& 7630\\

\enddata 
\tablenotetext{a}{The pointing center of the lead field is given; each is accompanied by a trail field 8~min later in R.A. Coordinates are J2000.}
\tablenotetext{b}{Data from this pointing were included in the deep dataset (Paper~II).}
\end{deluxetable*}

\end{document}